\documentclass[aps,prd,twocolumn,showpacs,10pt,superscriptaddress,preprintnumbers,nofootinbib]{revtex4-1}
\pdfoutput=1
\usepackage{graphicx}
\usepackage{subfigure}
\usepackage[dvipsnames]{xcolor}
\usepackage{amsmath}
\usepackage{mathrsfs}
\allowdisplaybreaks
\usepackage{amssymb}

\usepackage{slashed}
\usepackage[utf8]{inputenc}
\usepackage[colorlinks=true,linkcolor=blue,bookmarksopen,bookmarksnumbered]{hyperref}
\usepackage{color}
\usepackage[normalem]{ulem}
\usepackage{soul}
\usepackage{units}
\usepackage{rotating}
\usepackage{hhline,multirow,tabularx}
\usepackage{bm}
\usepackage{hyperref}
\usepackage[export]{adjustbox}
\usepackage{mdframed}
\usepackage{comment}
\usepackage{natbib}
\usepackage[absolute]{textpos}

\def\bea#1\eea{\begin{align}#1\end{align}} 

\newcommand{\bef}{\begin{figure}[htb]\centering}
\newcommand{\eef}{\end{figure}}

\newcommand{\nn}{\nonumber}

\newcommand{\shao}[1]{\marginpar{\footnotesize\textbf{SHAO}}}

\def\<{\langle}
\def\>{\rangle}

\def\cos{\hbox{cos}}
\def\sin{\hbox{sin}}

\allowdisplaybreaks

\usepackage{lipsum}

\begin{document}

\begin{textblock}{4}(0.2,0.1)   
  \includegraphics[width=2cm]{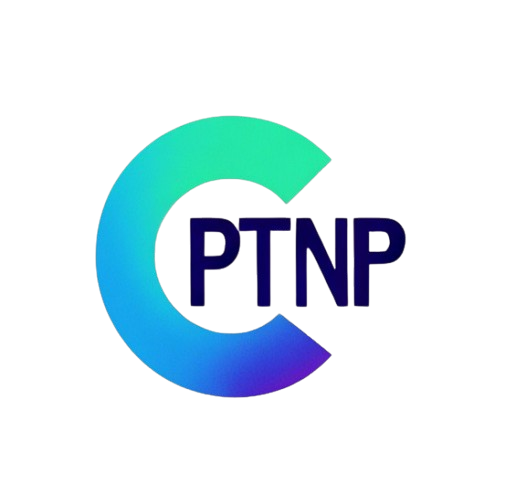}  
\end{textblock}

\begin{textblock}{6}(9,0.5)
  \raggedleft
  {\text{CPTNP-2025-008}}  
\end{textblock}

\title{Linearly Polarized Photon Fusion as a Precision Probe of the Tau Lepton Dipole Moments at Lepton Colliders}
	
\author{Ding Yu Shao}
\email{dingyu.shao@cern.ch}
\affiliation{Department of Physics, Center for Field Theory and Particle Physics, Fudan University, Shanghai, 200433, China}
\affiliation{Key Laboratory of
Nuclear Physics and Ion-beam Application (MOE), Fudan University, Shanghai, 200433, China}
\affiliation{Shanghai Research Center for Theoretical Nuclear Physics, NSFC and Fudan University, Shanghai 200438, China}
\affiliation{Center for High Energy Physics, Peking University, Beijing 100871, China}

\author{Hao Xiang}
\affiliation{Department of Physics, Center for Field Theory and Particle Physics, Fudan University, Shanghai, 200433, China}

\author{Fang Xu}
\email{xufang@wustl.edu}
\affiliation{Department of Physics, Center for Field Theory and Particle Physics, Fudan University, Shanghai, 200433, China}

\author{Bin Yan}
\email{yanbin@ihep.ac.cn}
\affiliation{Institute of High Energy Physics, Chinese Academy of Sciences, Beijing 100049, China}
\affiliation{Center for High Energy Physics, Peking University, Beijing 100871, China}

\author{Cheng Zhang}
\email{chengzhang@hznu.edu.cn}
\affiliation{School of Physics, Hangzhou Normal University, Hangzhou, Zhejiang 311121, China}


\begin{abstract}

We present a comprehensive investigation into the anomalous magnetic dipole moment ($a_\tau$) and electric dipole moment ($d_\tau$) of the $\tau$ lepton using the $\gamma\gamma \to \tau^+\tau^-$ process at future lepton colliders, with the Super Tau-Charm Facility serving as a benchmark. By employing transverse-momentum-dependent factorization, we introduce novel observables derived from $\cos2\phi$, $\sin2\phi$, and $\cos4\phi$ azimuthal asymmetries to precisely probe the $\tau$ lepton's electromagnetic structure. Our analysis significantly enhances the precision of $a_\tau$ constraints within the photon-photon fusion process, yielding $\mathrm{Re}(a_\tau) \in [-4.6, 7.0] \times 10^{-3}$ at the $2\sigma$ confidence level, which approaches the precision of the Standard Model prediction. These findings highlight the considerable potential of azimuthal asymmetry measurements for high-precision determinations of fundamental particle properties at future lepton colliders.

\end{abstract}

\maketitle

\section{Introduction}\label{sec:intro}

The anomalous magnetic dipole moments (MDM) and electric dipole moments (EDM) of leptons are fundamental probes of the Standard Model (SM) and new physics (NP)~\cite{Czarnecki:2001pv, Giudice:2012ms, Kurz:2014wya, Kurz:2015bia, Kurz:2016bau, Liu:2018xkx, Liu:2020qgx, Aebischer:2021uvt, Li:2021koa, Cirigliano:2021peb, BhupalDev:2021ipu, Afik:2022vpm, Xu:2023ene, Wen:2023xxc, Cao:2023juc, ParticleDataGroup:2024cfk}. The charge conjugation and parity ($CP$)-even MDM tests SM radiative corrections and the potential new heavy particles via quantum loops, whereas the $CP$-odd EDM is a particularly clean probe of new sources of $CP$ violation. The sensitivity of these dipole moments to NP is expected to be enhanced for heavier leptons~\cite{Giudice:2012ms}, rendering the massive $\tau$ lepton a promising probe of such effects. Recent results from the Fermilab Muon $g-2$ Collaboration, which have achieved unprecedented precision~\cite{Muong-2:2023cdq, Muong-2:2025xyk}, underscore the importance of these studies. Although recent theoretical calculations have reduced the longstanding tension between the measured muon MDM and its SM prediction~\cite{Aliberti:2025beg}, this development reinforces the need for precision studies of the $\tau$ lepton. Investigations of the $\tau$ dipole moments thus provide a complementary, and potentially flavor-dependent, window into NP.

The measurement of the $\tau$ lepton's dipole moments presents significant challenges due to its short lifetime, which prevents the direct methods employed for the electron and muon. Instead, these fundamental parameters must be inferred from the precise analysis of $\tau$-pair production and decay kinematics in high-energy collider experiments~\cite{delAguila:1991rm, DELPHI:2003nah, Bernabeu:2007rr, Atag:2010ja, Billur:2013rva, Eidelman:2016aih, Chen:2018cxt, Fu:2019utm, Beresford:2019gww, Dyndal:2020yen, Belle:2021ybo, Crivellin:2021spu, USBelleIIGroup:2022qro, ATLAS:2022ryk, CMS:2022arf, Verducci:2023cgx, Denizli:2024uwv, CMS:2024qjo, Gogniat:2025eom}. Recently, the ultraperipheral heavy ion collisions (UPCs) have emerged as a powerful, complementary approach for constraining these dipole moments, with two principal advantages:  (i) the coherent $Z^2$-enhanced photon flux and clean electromagnetic environment enable precise dipole moment measurements competitive with those obtained at conventional $e^+e^-$ or hadron colliders, as demonstrated by ATLAS and CMS results~\cite{Baltz:2007kq, ATLAS:2022ryk, CMS:2022arf}; (ii) the highly linear polarization of coherent photons induces an azimuthal angular asymmetry in $\tau$-pair production that provides simultaneous sensitivity to both the MDM and EDM effects~\cite{Li:2019yzy, Shao:2023bga}.

\bigskip

\begin{figure}[ht!]
    \centering
    \includegraphics[width=0.5\linewidth]{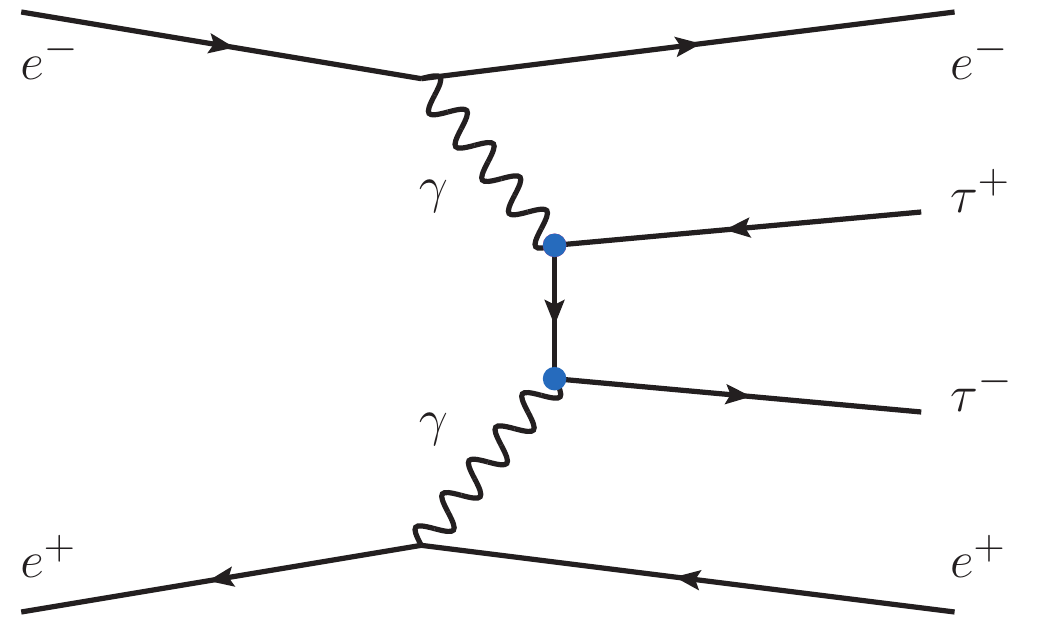}
    \caption{Production of a $\tau^+ \tau^-$ pair via photon-photon fusion at $e^+ e^-$ colliders. The blue dots represent the effective $\gamma \tau^+ \tau^-$ vertices.}
    \label{fig:diagram}
\end{figure}

While UPCs provide a powerful platform, their final sensitivity is limited by the uncertainties associated with the photon flux probability function, which depends on the modeling of the nuclear charge density distribution~\cite{Bertulani:1987tz, Vidovic:1992ik, Wang:2022ihj}. To overcome these limitations, we extend previous analyses of $\tau$ lepton dipole moments in UPCs from Ref.~\cite{Shao:2023bga} and explore the potential for measuring these parameters via the $\gamma \gamma \to \tau^+ \tau^-$ process at lepton colliders, as illustrated in Fig.~\ref{fig:diagram}. Promising venues include the proposed Super Tau-Charm Facility (STCF)~\cite{Achasov:2023gey} or Super Charm-Tau Factory~\cite{Barnyakov:2020vob}, designed specifically for precision $\tau$ physics. At such facilities, the photon flux from the leptons is calculable in perturbative QED, thereby mitigating the large theoretical uncertainties associated with the nuclear charge density in UPCs.

The electromagnetic properties of the $\tau$ lepton could be characterized by the effective $\gamma \tau^+ \tau^-$ vertex,
\begin{align}
\Gamma^\mu&(q^2)= \label{eq:vertex}\\
& -i e \Big\{ F_1(q^2) \gamma^\mu + \left[ i F_2(q^2) + F_3(q^2) \gamma_5 \right] \frac{\sigma^{\mu \nu} q_\nu}{2 m_\tau} \Big\} , \notag
\end{align}
where $\sigma^{\mu \nu}=i\left[\gamma^\mu, \gamma^\nu\right]/2$, $m_\tau$ represents the mass of the $\tau$ lepton, and $q$ denotes the momentum of the photon. In this analysis, we restrict ourselves to possible new-physics effects in the $\gamma \tau^+ \tau^-$ vertex, parametrized by the dipole moments, and neglect potential $\gamma \gamma \tau^+ \tau^-$ contact interactions, which correspond to higher-dimensional operators in effective field theory and are expected to be subleading. In the static limit of $q^\mu \to 0$, define the electric charge $F_1(0)=1$, the anomalous MDM $a_\tau = F_2(0)$, and the EDM $d_\tau = e F_3(0)/(2m_\tau)$. The SM predicts $a_\tau = 0.00117721(5)$~\cite{Eidelman:2007sb, ParticleDataGroup:2024cfk}, dominated by the one-loop Schwinger correction $\alpha_e/2\pi\simeq 0.00116$~\cite{Schwinger:1948iu}, while $d_\tau \sim \mathcal{O}(10^{-37}) \ e\,\mathrm{cm}$~\cite{Yamaguchi:2020eub,Yamaguchi:2020dsy} is negligible. The CMS Collaboration has recently established the most stringent constraint on the $\tau$ MDM, which yields $a_\tau = 0.0009^{+0.0032}_{-0.0031}$~\cite{CMS:2024qjo}, based on $\gamma\gamma \to \tau^+\tau^-$ production in proton-proton collisions. A notable earlier measurement came from
the DELPHI Collaboration, which obtained $a_\tau = -0.018 \pm 0.017$~\cite{DELPHI:2003nah} using $\gamma\gamma \to \tau^+\tau^-$ production in $e^+e^-$ collisions. While less precise than the CMS result, the DELPHI analysis provides unique advantages: it directly probes the $\gamma \tau^+ \tau^-$ vertex through experimental observables with minimal theoretical assumptions, such as those related to hadronic initial states or photon flux modeling inherent in hadron colliders. This approach benefits from the clean initial state of elementary particles and a well-characterized photon flux, providing complementary information to hadron collider measurements. For the EDM, the Belle Collaboration has set the tightest constraints, measuring $\mathrm{Re}(d_\tau) = (-0.62 \pm 0.63) \times 10^{-17} \ e\,\mathrm{cm}$ and $\mathrm{Im}(d_\tau) = (-0.40 \pm 0.32) \times 10^{-17} \ e\,\mathrm{cm}$. Given that the theoretical value of $d_\tau$ is currently far below the sensitivity of experimental measurements, we consider the SM values of $a_\tau \approx 0.00118$ and $d_\tau = 0$ in the analysis, with any observed deviation attributed to potential NP contributions.

Similar to UPCs, the lepton pair production via two-photon collision at $e^+e^-$ colliders can be described using the equivalent photon approximation (EPA)~\cite{vonWeizsacker:1934nji, Williams:1933, Williams:1934ad, Krauss:1997vr}, also known as the Weizsäcker-Williams method. The quasireal photons emitted from unpolarized electrons and positrons exhibit significant linear polarization, which can induce an azimuthal asymmetry in the final-state lepton distribution analogous to that observed in UPCs~\cite{STAR:2019wlg}. Crucially, the quasireal nature of the interacting photons ($q^2\approx 0$) ensures that this process directly probes the static dipole moments defined in Eq.~\eqref{eq:vertex}, in contrast to the high-$q^2$ process $e^+e^-\to \tau^+\tau^-$.

\section{Theoretical Formalism}\label{sec:theoretical}

We consider the production of $\tau^+\tau^-$ pairs in photon-photon collisions (Fig.~\ref{fig:diagram}),
\begin{align}
    \gamma(k_1)+\gamma(k_2) \to \tau^+(p_1) + \tau^-(p_2),
\end{align}
where the quasireal photons carry momenta $k_1$ and $k_2$. The dominant contribution to the azimuthal asymmetry occurs in the correlation limit $|\bm{q}_{\perp}| \ll |\bm{P}_{\perp}|$~\cite{Li:2019yzy, Li:2019sin, STAR:2019wlg, Xiao:2020ddm, Klein:2020jom, Zhao:2022dac, Brandenburg:2022tna, Zha:2018tlq, Wang:2021kxm, Wang:2022gkd}, with $\bm{P}_{\perp} \equiv (\bm{p}_{1 \perp} - \bm{p}_{2 \perp}) / 2$ and $\bm{q}_{\perp} \equiv \bm{p}_{1 \perp} + \bm{p}_{2 \perp} = \bm{k}_{1 \perp} + \bm{k}_{2 \perp}$. In this kinematic region, the $\tau^+\tau^-$ pair are produced nearly back to back, satisfying $\bm{P}_{\perp} \approx \bm{p}_{1 \perp} \approx -\bm{p}_{2 \perp}$, and the azimuthal asymmetry is characterized by the angle $\phi$ between the transverse momenta $\bm{q}_{\perp}$ and $\bm{P}_{\perp}$.

A systematic description of the azimuthal asymmetry in the kinematic region where $|\bm{q}_{\perp}| \ll |\bm{P}_{\perp}|$ is provided by the transverse-momentum-dependent (TMD) factorization framework. Analogous to the definition for gluon TMDs~\cite{Boussarie:2023izj}, the photon TMD distribution in an electron is defined by 
\begin{align}
    & f_{\gamma}^{\alpha \beta}(x,\bm{k}_{\perp})  \notag \\
   = & \int \frac{\mathrm{d} b^- \mathrm{d}^2 \bm{b}_\perp}{P^+ (2\pi)^3}e^{-i b^- (x P^+)+i \bm{b}_\perp\cdot \bm{k}_\perp} \notag \\
   &\hspace{1.5cm}\times\left\langle e(P)\left|F^{+\alpha}(b) F^{+\beta}(0)\right| e(P)\right\rangle \big|_{b^+=0}\notag \\
   = & \ -\frac{g_{\perp}^{\alpha \beta}}{2} x f(x,\bm{k}_{\perp}^2) + \left(\frac{g_{\perp}^{\alpha \beta}}{2} + \frac{k_{\perp}^\alpha k_{\perp}^\beta}{\bm{k}_{\perp}^2} \right) x h_1^{\perp}(x,\bm{k}_{\perp}^2),
\end{align}
where $x$ is the photon's longitudinal momentum fraction and $\bm{k}_\perp$ is its transverse momentum. The transverse metric tensor is $g_{\perp}^{\alpha \beta} \equiv g^{\alpha \beta} - (n_a^\alpha n_b^\beta + n_a^\beta n_b^\alpha)$ with light cone vectors $n_{a,b}^\mu = (1,0,0,\pm1)/\sqrt{2}$ and $b^\pm=(b^0\pm b^3)/\sqrt{2}$. The functions $f(x,\bm{k}_\perp^2)$ and $h_1^\perp(x,\bm{k}_\perp^2)$ correspond to the unpolarized and linearly polarized photon TMD distributions, respectively.

In contrast to UPCs, where the TMDs are nonperturbative and model dependent, a key advantage here is that these distributions are perturbatively calculable in QED. At leading order (LO), they are given by~\cite{Pisano:2013cya, Jia:2024xzx}
\begin{align}
f\left(x, \bm{k}_{\perp}^2\right) & =\frac{\alpha_e}{2 \pi^2} \frac{1+(1-x)^2}{x} \frac{\bm{k}_{\perp}^2}{\left(\bm{k}_{\perp}^2+x^2 m_e^2\right)^2}, \\
h_1^{\perp}\left(x, \bm{k}_{\perp}^2\right) & =\frac{\alpha_e}{\pi^2} \frac{1-x}{x} \frac{\bm{k}_{\perp}^2}{\left(\bm{k}_{\perp}^2+x^2 m_e^2\right)^2},
\end{align}
where $m_e$ is the electron mass and $\alpha_e$ is the fine-structure constant.

Within the TMD formalism~\cite{Boussarie:2023izj}, the differential cross section for the process  $e^+ e^- \to e^+ e^- + \tau^+ \tau^-$ can be expressed as follows:
\begin{align}
& \frac{\mathrm{d} \sigma}{\mathrm{d}^2 \bm{P}_{\perp} \mathrm{d}^2 \bm{q}_{\perp} \mathrm{d} y_1 \mathrm{d} y_2} = \frac{1}{64 \pi^2 s^2} \int \mathrm{d}^2 \bm{k}_{1 \perp} \mathrm{d}^2 \bm{k}_{2 \perp} \notag \\
&\ \ \times \delta^2\left(\bm{q}_{\perp}-\bm{k}_{1 \perp}-\bm{k}_{2 \perp}\right) x_1 x_2 \Big\{ \notag\\
&\ \ \left(\left|\mathcal{M}_{++}\right|^2+\left|\mathcal{M}_{--}\right|^2+\left|\mathcal{M}_{+-}\right|^2+\left|\mathcal{M}_{-+}\right|^2\right) \notag \\
&\ \ \quad\times f\left(x_1, \bm{k}_{1 \perp}^2\right) f\left(x_2, \bm{k}_{2 \perp}^2\right) \notag \\
&\ \ - 2 \mathrm{Re}\left[ \mathrm{e}^{2i(\phi_1-\phi_P)} \left( \mathcal{M}_{--} \mathcal{M}_{+-}^* + \mathcal{M}_{-+} \mathcal{M}_{++}^* \right)\right] \notag \\
&\ \ \quad\times h_1^{\perp}\left(x_1, \bm{k}_{1 \perp}^2\right) f\left(x_2, \bm{k}_{2 \perp}^2\right) \notag \\
&\ \ - 2 \mathrm{Re}\left[ \mathrm{e}^{2i(\phi_2-\phi_P)} \left( \mathcal{M}_{++} \mathcal{M}_{+-}^* + \mathcal{M}_{-+} \mathcal{M}_{--}^* \right)\right] \notag \\
&\ \ \quad\times f\left(x_1, \bm{k}_{1 \perp}^2\right) h_1^{\perp}\left(x_2, \bm{k}_{2 \perp}^2\right) \notag \\
&\ \ + 2 \mathrm{Re}\left[ \mathrm{e}^{2i(\phi_1 - \phi_2)} \left( \mathcal{M}_{--} \mathcal{M}_{++}^* \right)\right] \notag \\
&\ \ \quad\times h_1^{\perp}\left(x_1, \bm{k}_{1 \perp}^2\right) h_1^{\perp}\left(x_2, \bm{k}_{2 \perp}^2\right) \notag \\
&\ \ + 2 \mathrm{Re}\left[ \mathrm{e}^{2i(\phi_1 + \phi_2 - 2\phi_P)} \left( \mathcal{M}_{-+} \mathcal{M}_{+-}^* \right)\right] \notag \\
&\ \ \quad\times h_1^{\perp}\left(x_1, \bm{k}_{1 \perp}^2\right) h_1^{\perp}\left(x_2, \bm{k}_{2 \perp}^2\right)\Big\} ,
\label{eq:xsection1}
\end{align}
where $y_1$ and $y_2$ represent the rapidities of $\tau^+$ and $\tau^-$, and $s \equiv (p_1 + p_2)^2 = (k_1 + k_2)^2 = \left( |\bm{P}_\perp|^2 + m_\tau^2 \right) \left[ 2+2\cosh{(y_1 - y_2)} \right]$ indicates the invariant mass of the $\tau^+ \tau^-$ pair with $m_\tau$ being the tau mass. In Eq.~\eqref{eq:xsection1}, $\phi_P$ and $\phi_i$ denote the azimuthal angle of $\bm{P}_{\perp}$ and $\bm{k}_{i \perp}$ with $i = 1,\ 2$, respectively, while $\mathcal{M}_{\lambda_1,\lambda_2}$ represents the helicity amplitude for the process $\gamma(k_1,\lambda_1) \gamma(k_2,\lambda_2) \to \tau^+(p_1) \tau^-(p_2)$ at $\phi_P=0$. The photon momentum fractions $x_{1,2}$ are determined from the kinematics
\begin{align}
x_{1,2}=\sqrt{\dfrac{|\bm{P}_\perp|^2+m_\tau^2}{S_{ee}}}\left(e^{\pm y_1}+e^{\pm y_2}\right),
\end{align}
where $\sqrt{S_{ee}}$ being the center-of-mass energy of the $e^+ e^-$ collider.

Since the anomalous form factors are small ($|F_{2,3}| \ll 1$), we expand the differential cross section up to $\mathcal{O}(|F_{2,3}|^2)$ in their real and imaginary parts. The differential cross section then takes the following form:
\begin{widetext}
\begin{align}
\frac{\mathrm{d} \sigma}{\mathrm{d}^2 \bm{P}_{\perp} \mathrm{d}^2 \bm{q}_{\perp} \mathrm{d} y_1 \mathrm{d} y_2}&\ =\ \frac{\alpha_e^2}{2 \pi^2 M^4}\Big[ \notag\\
&\big( C_0 + C_0^{\mathrm{Re}(F_2)} \mathrm{Re}(F_2) + C_0^{\mathrm{Re}(F_2)^2} \mathrm{Re}(F_2)^2 + C_0^{\mathrm{Re}(F_3)^2} \mathrm{Re}(F_3)^2 + C_0^{\mathrm{Im}(F_2)^2} \mathrm{Im}(F_2)^2 + C_0^{\mathrm{Im}(F_3)^2} \mathrm{Im}(F_3)^2 \big) \notag\\
+& \big( C_{c2\phi} + C_{c2\phi}^{\mathrm{Re}(F_2)^2} \mathrm{Re}(F_2)^2 + C_{c2\phi}^{\mathrm{Re}(F_3)^2} \mathrm{Re}(F_3)^2 + C_{c2\phi}^{\mathrm{Im}(F_2)^2} \mathrm{Im}(F_2)^2 + C_{c2\phi}^{\mathrm{Im}(F_3)^2} \mathrm{Im}(F_3)^2 \big) \cos(2\phi) \notag\\
+& C_{s2\phi}^{\mathrm{Im}(F_2)\mathrm{Im}(F_3)}\mathrm{Im}(F_2)\mathrm{Im}(F_3)\sin(2\phi) \notag\\
+& \big( C_{c4\phi} + C_{c4\phi}^{\mathrm{Im}(F_2)^2} \mathrm{Im}(F_2)^2 + C_{c4\phi}^{\mathrm{Im}(F_3)^2} \mathrm{Im}(F_3)^2 \big) \cos(4\phi) \Big] ,
\label{eq:xsection3}
\end{align}
\end{widetext}
where $M=\sqrt{s}$, and $\phi \equiv \phi_q - \phi_P$ denotes the azimuthal angle between the transverse momentum $\bm{q}_{\perp}$ and $\bm{P}_{\perp}$. In this expansion, the form factors are treated as generally complex.

The angular structure of Eq.~\eqref{eq:xsection3} reveals a powerful, phase-sensitive probe of the dipole moments. Notably, the $\sin(2\phi)$ modulation arises exclusively from the interference of the imaginary parts of the form factors. Furthermore, these imaginary components generate new, distinct contributions to the $\cos(4\phi)$ asymmetry. This rich angular dependence is crucial for disentangling the real and imaginary components of $F_2$ and $F_3$ through measurements of the various azimuthal modulations. The coefficients $C_i$ appearing in Eq.~\eqref{eq:xsection3} can be expressed as follows:

\begin{align}
& C_0 = \frac{4 m_\tau^4}{(\bm{P}_{\perp}^2 + m_\tau^2)^2} \int \big[ \mathrm{d} \mathcal{K}_\perp \big] \Big [\big( -2 + \frac{M^2}{m_\tau^2} +\frac{\bm{P}_{\perp}^2 M^2}{m_\tau^4} \notag\\
& \: -2\frac{\bm{P}_{\perp}^4}{m_\tau^4} \big) x_1 f(x_1,\bm{k}_{1 \perp}^2) x_2 f(x_2,\bm{k}_{2 \perp}^2) \notag\\
& \: -2 x_1 h_1^\perp(x_1,\bm{k}_{1 \perp}^2) x_2 h_1^\perp(x_2,\bm{k}_{2 \perp}^2) \cos(2\phi_1 - 2\phi_2) \Big] ,
\label{eq:A0} \\[7pt]
& C_{c2\phi} = \frac{16 \bm{P}_{\perp}^2 m_\tau^2}{(\bm{P}_{\perp}^2 + m_\tau^2)^2} \int \big[ \mathrm{d} \mathcal{K}_\perp \big] \notag\\
& \: \Big[ x_1 h_1^\perp(x_1,\bm{k}_{1 \perp}^2) x_2 f(x_2,\bm{k}_{2 \perp}^2) \cos(2\phi_1 - 2\phi_q) \notag\\
& \: + x_1 f(x_1,\bm{k}_{1 \perp}^2) x_2 h_1^\perp(x_2,\bm{k}_{2 \perp}^2) \cos(2\phi_2 - 2\phi_q) \Big] ,
\label{eq:A2} \\[7pt]
& C_{c4\phi} = \frac{-8 \bm{P}_{\perp}^4}{(\bm{P}_{\perp}^2 + m_\tau^2)^2} \int \big[ \mathrm{d} \mathcal{K}_\perp \big] \notag\\
& \: x_1 h_1^\perp(x_1,\bm{k}_{1 \perp}^2) x_2 h_1^\perp(x_2,\bm{k}_{2 \perp}^2) \cos(2\phi_1 + 2\phi_2 - 4\phi_q) ,
\label{eq:A4} \\
& C_0^{\mathrm{Re}(F_2)} = \frac{8 M^2}{\bm{P}_{\perp}^2 + m_\tau^2} \int \big[ \mathrm{d} \mathcal{K}_\perp \big] \Big[ x_1 f(x_1,\bm{k}_{1 \perp}^2) x_2 f(x_2,\bm{k}_{2 \perp}^2) \notag\\
& \: - x_1 h_1^\perp(x_1,\bm{k}_{1 \perp}^2) x_2 h_1^\perp(x_2,\bm{k}_{2 \perp}^2) \cos(2\phi_1 - 2\phi_2) \Big] ,
\label{eq:B0(1)} \\[7pt]
& C_0^{\mathrm{Re}(F_2)^2} = \frac{2 M^2}{\bm{P}_{\perp}^2 + m_\tau^2} \int \big[ \mathrm{d} \mathcal{K}_\perp \big] \notag\\
& \: \Big[ \big( 5 + 4\frac{\bm{P}_{\perp}^2}{m_\tau^2} \big) x_1 f(x_1,\bm{k}_{1 \perp}^2) x_2 f(x_2,\bm{k}_{2 \perp}^2) \notag\\
& \: - 5 x_1 h_1^\perp(x_1,\bm{k}_{1 \perp}^2) x_2 h_1^\perp(x_2,\bm{k}_{2 \perp}^2) \cos(2\phi_1 - 2\phi_2) \Big] ,
\label{eq:B0(2)} \\[7pt]
& C_0^{\mathrm{Re}(F_3)^2} = \frac{2 M^2}{\bm{P}_{\perp}^2 + m_\tau^2} \int \big[ \mathrm{d} \mathcal{K}_\perp \big] \notag\\
& \: \Big[ \big(3 + 4\frac{\bm{P}_{\perp}^2}{m_\tau^2} \big) x_1 f(x_1,\bm{k}_{1 \perp}^2) x_2 f(x_2,\bm{k}_{2 \perp}^2) \notag\\
& \: + 5 x_1 h_1^\perp(x_1,\bm{k}_{1 \perp}^2) x_2 h_1^\perp(x_2,\bm{k}_{2 \perp}^2) \cos(2\phi_1 - 2\phi_2) \Big] ,
\label{eq:C0(2)} \\[7pt]
& C_{c2\phi}^{\mathrm{Re}(F_2)^2} = C_{c2\phi}^{\mathrm{Re}(F_3)^2} = \frac{-4 \bm{P}_{\perp}^2 M^2}{m_\tau^2 (\bm{P}_{\perp}^2 + m_\tau^2)} \int \big[ \mathrm{d} \mathcal{K}_\perp \big] \notag\\
& \: \Big[ x_1 h_1^\perp(x_1,\bm{k}_{1 \perp}^2) x_2 f(x_2,\bm{k}_{2 \perp}^2) \cos(2\phi_1 - 2\phi_q) \notag\\
& \: + x_1 f(x_1,\bm{k}_{1 \perp}^2) x_2 h_1^\perp(x_2,\bm{k}_{2 \perp}^2) \cos(2\phi_2 - 2\phi_q) \Big] ,
\label{eq:B2(2)C2(2)} \\[7pt]
& C_0^{\mathrm{Im}(F_2)^2} = \frac{2 M^2 m_\tau^2}{(\bm{P}_{\perp}^2 + m_\tau^2)^2} \int \big[ \mathrm{d} \mathcal{K}_\perp \big] \notag\\
& \: \Big[ \big(3 + 5\frac{\bm{P}_{\perp}^2}{m_\tau^2} + 2\frac{\bm{P}_{\perp}^4}{m_\tau^4} \big) x_1 f(x_1,\bm{k}_{1 \perp}^2) x_2 f(x_2,\bm{k}_{2 \perp}^2) - \big(3 \notag\\
& \: + \frac{\bm{P}_{\perp}^2}{m_\tau^2} \big) x_1 h_1^\perp(x_1,\bm{k}_{1 \perp}^2) x_2 h_1^\perp(x_2,\bm{k}_{2 \perp}^2) \cos(2\phi_1 - 2\phi_2) \Big] ,
\label{eq:D0(2)} \\
& C_0^{\mathrm{Im}(F_3)^2} = \frac{2 M^2 m_\tau^2}{(\bm{P}_{\perp}^2 + m_\tau^2)^2} \int \big[ \mathrm{d} \mathcal{K}_\perp \big] \notag\\
& \: \Big[ \big(5 + 3\frac{\bm{P}_{\perp}^2}{m_\tau^2} + 2\frac{\bm{P}_{\perp}^4}{m_\tau^4} \big) x_1 f(x_1,\bm{k}_{1 \perp}^2) x_2 f(x_2,\bm{k}_{2 \perp}^2) + \big(3 \notag\\
& \: + \frac{\bm{P}_{\perp}^2}{m_\tau^2} \big) x_1 h_1^\perp(x_1,\bm{k}_{1 \perp}^2) x_2 h_1^\perp(x_2,\bm{k}_{2 \perp}^2) \cos(2\phi_1 - 2\phi_2) \Big] ,
\label{eq:E0(2)} \\
& C_{c2\phi}^{\mathrm{Im}(F_2)^2} = \frac{-4 \bm{P}_{\perp}^2 M^2}{m_\tau^2 (\bm{P}_{\perp}^2 + m_\tau^2)} \int \big[ \mathrm{d} \mathcal{K}_\perp \big] \notag\\
& \: \Big[ x_1 h_1^\perp(x_1,\bm{k}_{1 \perp}^2) x_2 f(x_2,\bm{k}_{2 \perp}^2) \cos(2\phi_1 - 2\phi_q) \notag\\
& \: + x_1 f(x_1,\bm{k}_{1 \perp}^2) x_2 h_1^\perp(x_2,\bm{k}_{2 \perp}^2) \cos(2\phi_2 - 2\phi_q) \Big] ,
\label{eq:D2(2)} \\[7pt]
& C_{c2\phi}^{\mathrm{Im}(F_3)^2} = \frac{-4 \bm{P}_{\perp}^2 M^2 (\bm{P}_{\perp}^2 + 3m_\tau^2)}{m_\tau^2 (\bm{P}_{\perp}^2 + m_\tau^2)^2} \int \big[ \mathrm{d} \mathcal{K}_\perp \big] \notag\\
& \: \Big[ x_1 h_1^\perp(x_1,\bm{k}_{1 \perp}^2) x_2 f(x_2,\bm{k}_{2 \perp}^2) \cos(2\phi_1 - 2\phi_q) \notag\\
& \: + x_1 f(x_1,\bm{k}_{1 \perp}^2) x_2 h_1^\perp(x_2,\bm{k}_{2 \perp}^2) \cos(2\phi_2 - 2\phi_q) \Big] ,
\label{eq:E2(2)} \\[7pt]
& C_{s2\phi}^{\mathrm{Im}(F_2)\mathrm{Im}(F_3)} = \frac{-8 \bm{P}_{\perp}^2 M^2}{(\bm{P}_{\perp}^2 + m_\tau^2)^2} \int \big[ \mathrm{d} \mathcal{K}_\perp \big] \notag\\
& \: \Big[ x_1 h_1^\perp(x_1,\bm{k}_{1 \perp}^2) x_2 f(x_2,\bm{k}_{2 \perp}^2) \cos(2\phi_1 - 2\phi_q) \notag\\
& \: - x_1 f(x_1,\bm{k}_{1 \perp}^2) x_2 h_1^\perp(x_2,\bm{k}_{2 \perp}^2) \cos(2\phi_2 - 2\phi_q) \Big] ,
\label{eq:G2(2)} \\[7pt]
& C_{c4\phi}^{\mathrm{Im}(F_2)^2} = C_{c4\phi}^{\mathrm{Im}(F_3)^2} = \frac{4 \bm{P}_{\perp}^4 M^2}{m_\tau^2 (\bm{P}_{\perp}^2 + m_\tau^2)^2} \int \big[ \mathrm{d} \mathcal{K}_\perp \big] \notag\\
& \: x_1 h_1^\perp(x_1,\bm{k}_{1 \perp}^2) x_2 h_1^\perp(x_2,\bm{k}_{2 \perp}^2) \cos(2\phi_1 + 2\phi_2 - 4\phi_q) ,
\label{eq:D4(2)E4(2)}
\end{align}
where the abbreviated notation for the integral is defined as shown:
\begin{align}
\int \big[ \mathrm{d} \mathcal{K}_\perp \big] \equiv \pi^2 \int \mathrm{d}^2 \bm{k}_{1 \perp} \mathrm{d}^2 \bm{k}_{2 \perp} \delta^2\left(\bm{q}_{\perp}-\bm{k}_{1 \perp}-\bm{k}_{2 \perp}\right).
\label{eq:abbreviated notation}
\end{align}

Here, the structure of these coefficients follows systematic patterns determined by the photon polarizations. The dependence on the individual photon azimuthal angles, $\phi_{1,2}$, is tied exclusively to the linearly polarized TMD $h_1^\perp$, reflecting its directional nature; the unpolarized TMD $f$ is azimuthally symmetric, introducing no $\phi_i$ dependence~\cite{Pisano:2013cya}.

This leads to a direct mapping between the initial polarization states and the final-state angular modulations in $\phi$. Specifically, configurations with one linearly polarized photon generate the $\cos(2\phi)$ and $\sin(2\phi)$ terms, while the $\cos(4\phi)$ modulation requires both photons to be linearly polarized\footnote{These characteristic azimuthal dependencies are analogous to those observed in transverse polarization phenomena of fermions~\cite{Wen:2023xxc, Wang:2024zns, Wen:2024cfu, Wen:2024nff, Cheng:2025cuv}.}. The azimuthally-independent part of the cross section receives contributions from initial states where both photons are unpolarized, and where both are polarized. Another source of azimuthal modulations arises from soft photon radiation beyond LO. The required QED resummation has been developed in Refs.~\cite{Hatta:2021jcd, Shao:2022stc, Shao:2023zge}. A detailed implementation of these high order corrections is left for future work.

\section{Phenomenology and Discussions}\label{sec:pheno}

\begin{figure*}[ht!]
\centering
{\includegraphics[width=0.5\textwidth]{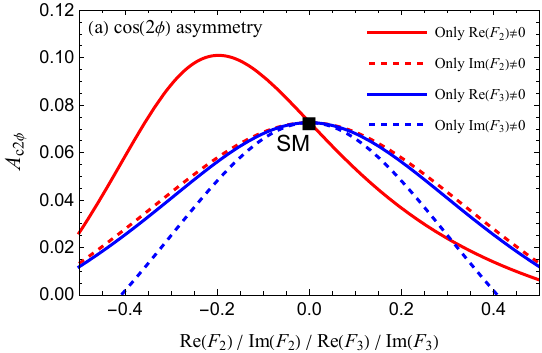}}
\hspace{0.2cm}
{ \includegraphics[width=0.4\textwidth]{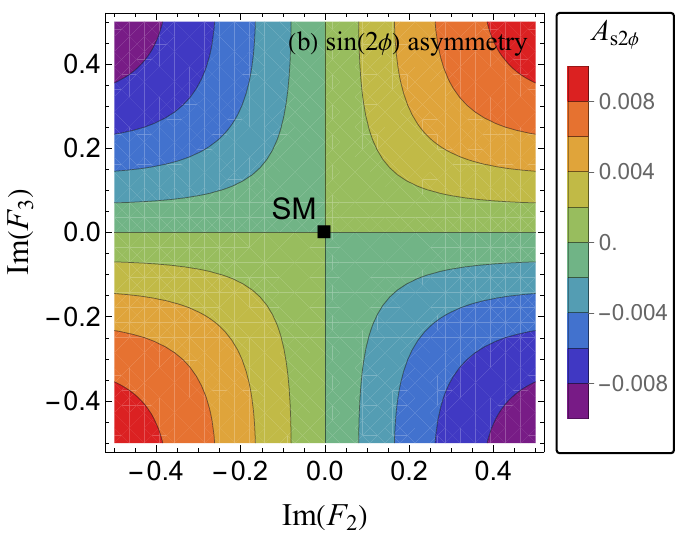}}
\\[20pt] 
{\includegraphics[width=0.5\textwidth]{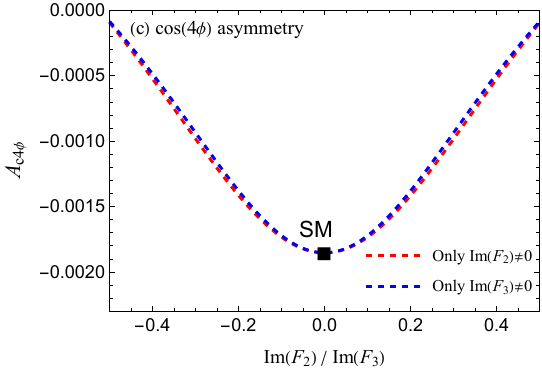}}
\caption{Theoretical predictions for (a) $\cos2\phi$, (b) $\sin2\phi$, and (c) $\cos4\phi$ azimuthal asymmetries from form factors $F_2$ and $F_3$ in $\gamma\gamma\to \tau^+ \tau^-$ at an $e^+ e^-$ collider with $\sqrt{S_{ee}} = 7$~GeV. The black square indicates the prediction from the SM.}
\label{fig:asymmetry}
\end{figure*}

To isolate the angular modulations derived in Eq.~\eqref{eq:xsection3}, we construct the following experimental asymmetries:
\begin{align}
A_{c 2 \phi}&=\frac{\sigma(\cos 2 \phi>0)-\sigma(\cos 2 \phi<0)}{\sigma(\cos 2 \phi>0)+\sigma(\cos 2 \phi<0)} ,\nn\\
A_{s2\phi}&=\frac{\sigma(y\times\sin 2 \phi>0)-\sigma(y\times\sin 2 \phi<0)}{\sigma(\sin 2 \phi>0)+\sigma(\sin 2 \phi<0)} ,\nn\\
A_{c 4\phi}&=\frac{\sigma(\cos 4 \phi>0)-\sigma(\cos 4 \phi<0)}{\sigma(\cos 4 \phi>0)+\sigma(\cos 4 \phi<0)} ,
\label{eq:Aphi}
\end{align}
where $\sigma(\mathcal{O}>0)$ denotes the cross section integrated over the phase space where the observable $\mathcal{O}$ is positive. The asymmetry $A_{s2\phi}$ is weighted by the $\tau$-pair rapidity, $y \equiv (y_1 + y_2)/2$. This weighting is necessary because the corresponding coefficient, $C_{s2\phi}^{\mathrm{Im}F_2\mathrm{Im}F_3}$, is antisymmetric under $y \to -y$ [cf. Eq.~\eqref{eq:G2(2)}], and this definition ensures a nonvanishing, $CP$-violating signal after integration over the full rapidity range.

The statistical uncertainties of these azimuthal asymmetries $A_i$ with $i=c2\phi,s2\phi,c4\phi$ are given by
\begin{align}
\delta A_i = \sqrt{\frac{1-A_i^2}{\sigma \cdot \mathscr{L}}} \simeq \frac{1}{\sqrt{\sigma \cdot \mathscr{L}}} ,
\label{eq:uncertainty}
\end{align}
where the approximation holds due to the small values of $A_i$ in the SM. A significant advantage of such ratio observables is the high degree of cancellation of systematic uncertainties (e.g., from luminosity and detector efficiencies). We therefore assume that the sensitivities are statistically limited and neglect systematic effects in the following analysis.

Building upon the above analysis, we provide the results for the production of $\tau^+ \tau^-$ pairs, using the STCF~\cite{Achasov:2023gey} as a benchmark machine. The center-of-mass energy, integrated luminosity, and kinematic cuts in this analysis are summarized as shown below:
\begin{align}
&\sqrt{S_{ee}} = 7\ \mathrm{GeV},\ \mathscr{L}_\mathrm{int} = 3\ \mathrm{ab}^{-1}, \notag \\
&\ \ \ \ \ 0.2\ \mathrm{GeV} < |\bm{P}_{\perp}| < 3\ \mathrm{GeV}, \notag \\
&\ \ \ |\bm{q}_{\perp}| < 0.1\ \mathrm{GeV},\ |y_{1,2}| < 0.5.
\label{eq:cuts}
\end{align}
The strict requirement on $|\bm{q}_{\perp}|$ ensures the applicability of the formalism in the correlation limit ($|\bm{q}_{\perp}| \ll |\bm{P}_{\perp}|$).
As a detailed STCF~\cite{Achasov:2023gey} detector simulation is not yet available, we simulate $\tau$ decays with TAUOLA~\cite{Davidson:2010rw} and fold acceptance $\times$ efficiency $(A\times\varepsilon)$ in our analysis. We adopt a tag-and-probe topology with one leptonic ($e$ or $\mu$) and one hadronic 1-prong $\tau$ decay and require the visible decay products to satisfy $p_T > 0.3$~GeV and $|\eta| < 2$. For representative working points, we use the per-object efficiencies $\varepsilon(e) = 0.95$, $\varepsilon(\mu) = 0.93$, and $\varepsilon(\tau_h,1\text{-pr}) = 0.85$. As a comparison, the implied overall efficiency of the process is $\sim20\%$, consistent in magnitude with Belle and Belle~II performance~\cite{Bodrov:2024wrw, BelleII:2020tracking, Banerjee:2022vdd}.

To optimize the measurement of the $CP$-violating asymmetry $A_{s2\phi}$, an additional selection of $|y|>0.4$ is imposed. This cut enhances the signal-to-background ratio, as the underlying coefficient of the $\sin(2\phi)$ modulation grows with $|y|$ relative to the azimuthally independent SM background.

\begin{figure*}[ht!]
\centering
{\includegraphics[width=0.405\textwidth]{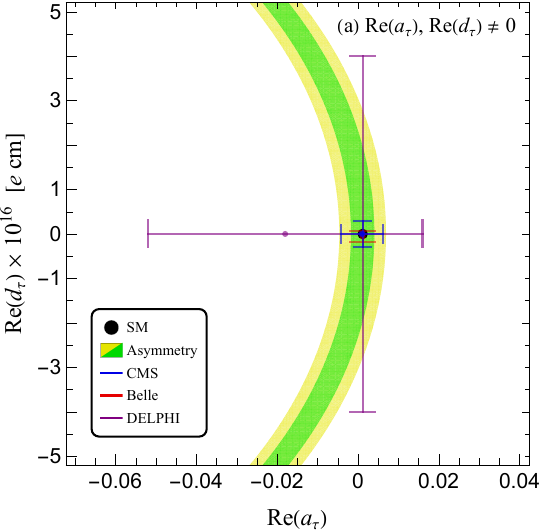}}
\hspace{3pt}
{\includegraphics[width=0.41\textwidth]{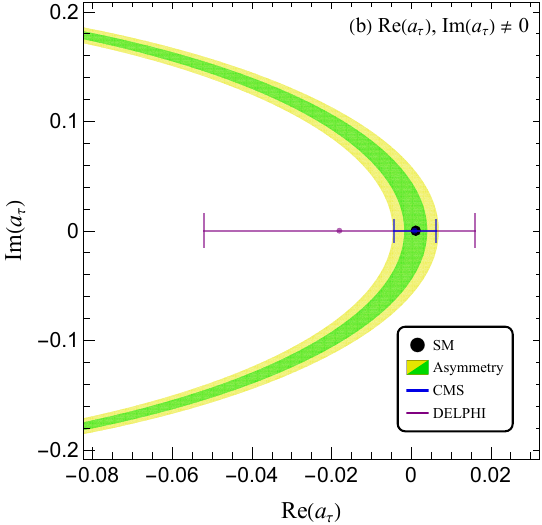}}
\\[10pt]
\hspace{5pt}{\includegraphics[width=0.4\textwidth]{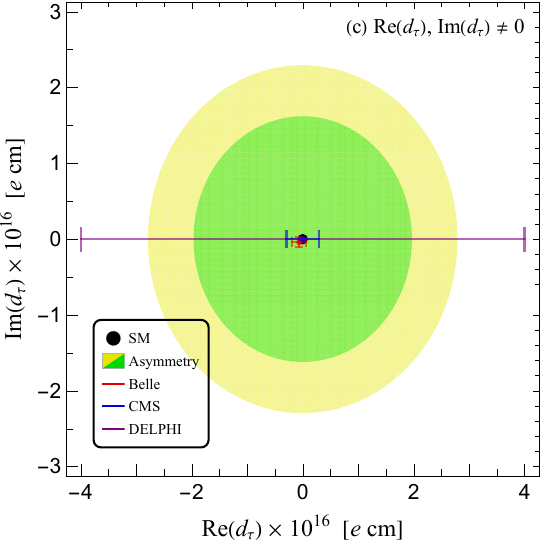}}
\hspace{11pt}
{\includegraphics[width=0.41\textwidth]{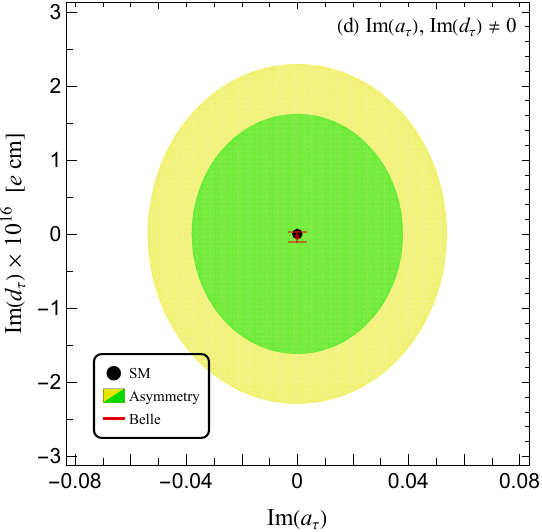}}
\caption{The expected sensitivity of the $\cos2\phi$ asymmetry to $\tau$ dipole moments for four benchmark scenarios: (a) $\mathrm{Re}(a_\tau), \mathrm{Re}(d_\tau)\neq 0$, (b) $\mathrm{Re}(a_\tau), \mathrm{Im}(a_\tau)\neq 0$, (c) $\mathrm{Re}(d_\tau), \mathrm{Im}(d_\tau)\neq 0$, (d) $\mathrm{Im}(a_\tau), \mathrm{Im}(d_\tau)\neq 0$. The green and yellow regions correspond to $1\sigma$ and $2\sigma$ CL, respectively. The black dots indicate SM predictions, while the blue (CMS)~\cite{CMS:2024qjo}, purple (DELPHI)~\cite{DELPHI:2003nah}, and red (Belle)~\cite{Belle:2021ybo} error bars represent current experimental constraints at $2\sigma$ CL.}
\label{fig:constraints}
\end{figure*}

Figure~\ref{fig:asymmetry} shows the theoretical predictions for the $\cos2\phi$, $\sin2\phi$ and $\cos4\phi$ asymmetries in $\gamma\gamma\to \tau^+\tau^-$ production at the STCF, using the kinematic cuts defined in Eq.~\eqref{eq:cuts}. The black squares represent the SM predictions. For the $A_{c2\phi}$ asymmetry, we examine individual form factor contributions, finding that they can significantly modify the asymmetry. In contrast, the $A_{s2\phi}$ and $A_{c4\phi}$ asymmetries are considerably smaller than $A_{c2\phi}$, and $A_{s2\phi}$ depends on a combination of the imaginary parts of $F_2$ and $F_3$, as shown in Eq.~\eqref{eq:xsection3}.

Figure~\ref{fig:constraints} displays the expected sensitivity of the $\cos2\phi$ azimuthal asymmetry to the $\tau$ lepton's anomalous magnetic and electric dipole moments, presenting four benchmark scenarios with $1\sigma$ (green) and $2\sigma$ (yellow) confidence-level (CL) regions that incorporate statistical uncertainties and the $\tau$-pair signal efficiency, where black dots indicate the SM predictions in each panel. These results are compared against existing $2\sigma$ constraints from CMS (blue error bars)~\cite{CMS:2024qjo}, DELPHI (purple error bars)~\cite{DELPHI:2003nah}, and Belle (red error bars)~\cite{Belle:2021ybo}. The analysis reveals $\mathrm{Re}(a_\tau)$ as the most experimentally accessible parameter, benefiting from its linear term with a large coefficient in Eq.~\eqref{eq:xsection3}. Under the assumptions $d_\tau = 0$ and $\mathrm{Im}(a_\tau) = 0$, we constrain $\mathrm{Re}(a_\tau)$ to
\begin{align}
-4.6 \times 10^{-3} < \mathrm{Re}(a_\tau) < 7.0 \times 10^{-3} ,
\label{eq:result_atau}
\end{align}
at $2\sigma$ CL. Our results demonstrate comparable sensitivity to the recent CMS measurement of $-4.2\times 10^{-3} < a_\tau < 6.2\times 10^{-3}$ at 95\% CL~\cite{CMS:2024qjo}, but without relying on the assumption about photon fluxes, and approaches the SM prediction of $a_\tau = 1.17721(5) \times 10^{-3}$~\cite{Eidelman:2007sb}. This result represents an order-of-magnitude improvement in precision compared to the earlier DELPHI measurement $a_\tau = -0.018 \pm 0.017$~\cite{DELPHI:2003nah}, demonstrating the potential of photon-fusion observables for high-precision studies in leptonic collision environments.

\begin{figure*}[ht!]
    \centering
    \includegraphics[width=0.4\textwidth]{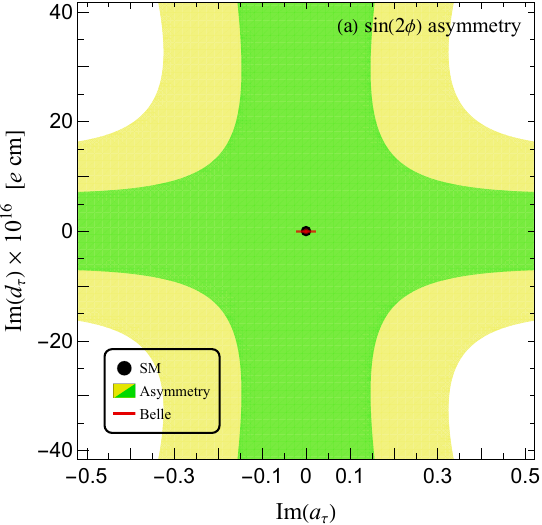}
    \hspace{10pt}
    \includegraphics[width=0.4\textwidth]{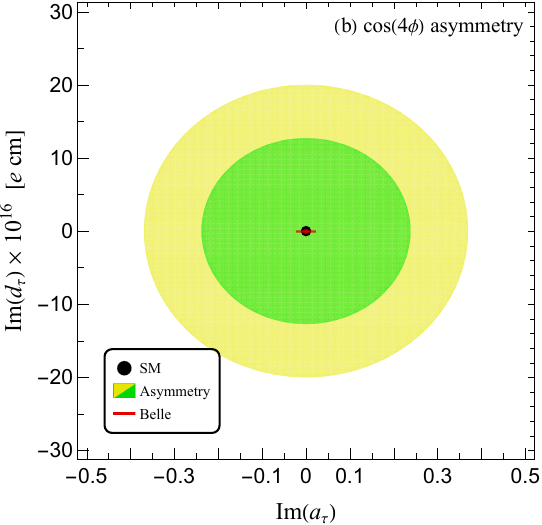}
    \caption{The expected sensitivity of the (a) $\sin2\phi$ and (b) $\cos4\phi$ asymmetries to $\tau$ dipole moments, with green and yellow regions correspond to $1\sigma$ and $2\sigma$ CL, respectively. The black dots indicate SM predictions, while the red error bars 
    indicate current experimental constraints at $2\sigma$ CL from Belle~\cite{Belle:2021ybo}.}
    \label{fig:constraints2}
\end{figure*}

However, our observable provides relatively weak constraints on $d_\tau$ due to $CP$ symmetry forbidding linear $d_\tau$ terms in the $\cos2\phi$ asymmetry. Under the assumptions $a_\tau = a_\tau^{\mathrm{SM}} \approx 1.18 \times 10^{-3}$ and $\mathrm{Im}(d_\tau) = 0$, we obtain the following constraint at $2\sigma$ CL:
\begin{align}
|\mathrm{Re}(d_\tau)| < 2.8 \times 10^{-16} \ e\,\mathrm{cm} .
\label{eq:result_dtau}
\end{align}
While the result represents a twofold improvement in precision over the DELPHI result~\cite{DELPHI:2003nah}, it remains approximately an order of magnitude less precise than Belle's measurement of $\mathrm{Re}(d_\tau) = (-0.62 \pm 0.63) \times 10^{-17} \ e\,\mathrm{cm}$ and than the CMS constraint of $|d_\tau|<2.9\times 10^{-17} \ e\,\mathrm{cm}$ (95\% CL), obtained under the single-operator assumption. The Belle Collaboration achieved this precision using the optimal observable method~\cite{Atwood:1991ka}, which constructs $CP$-odd spin-momentum correlations that depend linearly on $\mathrm{Re}(d_\tau)$ or $\mathrm{Im}(d_\tau)$~\cite{Belle:2002nla,Belle:2021ybo}. Future studies of the $\gamma\gamma\to\tau^+\tau^-$ process could substantially enhance EDM sensitivity through the implementation of similar $CP$-violating observables~\cite{Huang:1996jr,Bernreuther:1996dr}, and we will investigate this approach in future work. It is worth noting that the Belle II experiment at the KEKB accelerator is expected to reach an even higher sensitivity, down to the $\mathcal{O}(10^{-20})\ e\,\mathrm{cm}$ level, with an integrated luminosity of $50\ \mathrm{ab}^{-1}$~\cite{Bernreuther:2021elu}.

Figure~\ref{fig:constraints2} displays the sensitivity of $\sin2\phi$ and $\cos4\phi$ azimuthal asymmetries to $\mathrm{Im}(a_\tau)$ and $\mathrm{Im}(d_\tau)$. While these asymmetries provide weaker constraints compared to other observables, they exhibit a unique and complementary sensitivity: the $\sin2\phi$ asymmetry depends linearly on the product $\mathrm{Im}(a_\tau)\cdot\mathrm{Im}(d_\tau)$, whereas the $\cos4\phi$ asymmetry shows a quadratic dependence on $\mathrm{Im}(a_\tau)^2$ and $\mathrm{Im}(d_\tau)^2$. Crucially, both asymmetries are completely independent of real components and thus serve as distinctive probes of potential $CP$-violating effects.

\section{Conclusions}\label{sec:conclusion}

In this study, we have proposed a new method to probe the MDMs and EDMs of the $\tau$ lepton via the process $\gamma \gamma \to \tau^+ \tau^-$ at $e^+e^-$ colliders. Using the TMD factorization formalism, we demonstrated that azimuthal asymmetries in the final state provide powerful probes of the dipole form factors. Our phenomenological analysis, based on the projected parameters of the STCF, shows that the $\cos(2\phi)$ asymmetry is particularly sensitive to the MDM. We project a $2\sigma$ constraint of $-4.6 \times 10^{-3} < \mathrm{Re}(a_\tau)<7.0 \times 10^{-3}$, reaching a level of precision comparable to the strongest existing experimental limits, but without relying on assumptions about photon fluxes. The same measurement would constrain the EDM to $|\mathrm{Re}(d_\tau)|< 2.8 \times 10^{-16} \ e\,\mathrm{cm}$ at the $2\sigma$ CL.

Furthermore, we identified the $\sin(2\phi)$ and $\cos(4\phi)$ asymmetries as unique and complementary observables. They are sensitive exclusively to the imaginary components of the form factors, providing a clean channel to search for new physics. This work establishes the study of azimuthal asymmetries in two-photon processes as a promising new avenue for precision $\tau$ physics and beyond-the-Standard-Model searches at future lepton colliders.

\vspace{3mm}
\noindent{\it Acknowledgments.}
The authors thank C.-P. Yuan for the helpful discussion, and Yu-Bo Li for bringing the CMS result to our attention. F.X. also thanks Yoav Afik for valuable discussions on $\tau$ efficiency. This work is supported by the National Science Foundations of China under Grants No.~12275052 and No.~12147101.
B.Y. is supported in part by the National Science Foundation of China under Grant No.~12422506, the IHEP under Grant No.~E25153U1 and CAS under Grant No.~E429A6M1. The authors gratefully acknowledge the valuable discussions and insights provided by the members of the China Collaboration of Precision Testing and New Physics.

\bibliography{ref}

\begin{thebibliography}{82}%
\makeatletter
\providecommand \@ifxundefined [1]{%
 \@ifx{#1\undefined}
}%
\providecommand \@ifnum [1]{%
 \ifnum #1\expandafter \@firstoftwo
 \else \expandafter \@secondoftwo
 \fi
}%
\providecommand \@ifx [1]{%
 \ifx #1\expandafter \@firstoftwo
 \else \expandafter \@secondoftwo
 \fi
}%
\providecommand \natexlab [1]{#1}%
\providecommand \enquote  [1]{``#1''}%
\providecommand \bibnamefont  [1]{#1}%
\providecommand \bibfnamefont [1]{#1}%
\providecommand \citenamefont [1]{#1}%
\providecommand \href@noop [0]{\@secondoftwo}%
\providecommand \href [0]{\begingroup \@sanitize@url \@href}%
\providecommand \@href[1]{\@@startlink{#1}\@@href}%
\providecommand \@@href[1]{\endgroup#1\@@endlink}%
\providecommand \@sanitize@url [0]{\catcode `\\12\catcode `\$12\catcode `\&12\catcode `\#12\catcode `\^12\catcode `\_12\catcode `\%12\relax}%
\providecommand \@@startlink[1]{}%
\providecommand \@@endlink[0]{}%
\providecommand \url  [0]{\begingroup\@sanitize@url \@url }%
\providecommand \@url [1]{\endgroup\@href {#1}{\urlprefix }}%
\providecommand \urlprefix  [0]{URL }%
\providecommand \Eprint [0]{\href }%
\providecommand \doibase [0]{http://dx.doi.org/}%
\providecommand \selectlanguage [0]{\@gobble}%
\providecommand \bibinfo  [0]{\@secondoftwo}%
\providecommand \bibfield  [0]{\@secondoftwo}%
\providecommand \translation [1]{[#1]}%
\providecommand \BibitemOpen [0]{}%
\providecommand \bibitemStop [0]{}%
\providecommand \bibitemNoStop [0]{.\EOS\space}%
\providecommand \EOS [0]{\spacefactor3000\relax}%
\providecommand \BibitemShut  [1]{\csname bibitem#1\endcsname}%
\let\auto@bib@innerbib\@empty
\bibitem [{\citenamefont {Czarnecki}\ and\ \citenamefont {Marciano}(2001)}]{Czarnecki:2001pv}%
  \BibitemOpen
  \bibfield  {author} {\bibinfo {author} {\bibfnamefont {A.}~\bibnamefont {Czarnecki}}\ and\ \bibinfo {author} {\bibfnamefont {W.~J.}\ \bibnamefont {Marciano}},\ }\href {\doibase 10.1103/PhysRevD.64.013014} {\bibfield  {journal} {\bibinfo  {journal} {Phys. Rev. D}\ }\textbf {\bibinfo {volume} {64}},\ \bibinfo {pages} {013014} (\bibinfo {year} {2001})},\ \Eprint {http://arxiv.org/abs/hep-ph/0102122} {arXiv:hep-ph/0102122} \BibitemShut {NoStop}%
\bibitem [{\citenamefont {Giudice}\ \emph {et~al.}(2012)\citenamefont {Giudice}, \citenamefont {Paradisi},\ and\ \citenamefont {Passera}}]{Giudice:2012ms}%
  \BibitemOpen
  \bibfield  {author} {\bibinfo {author} {\bibfnamefont {G.~F.}\ \bibnamefont {Giudice}}, \bibinfo {author} {\bibfnamefont {P.}~\bibnamefont {Paradisi}}, \ and\ \bibinfo {author} {\bibfnamefont {M.}~\bibnamefont {Passera}},\ }\href {\doibase 10.1007/JHEP11(2012)113} {\bibfield  {journal} {\bibinfo  {journal} {JHEP}\ }\textbf {\bibinfo {volume} {11}},\ \bibinfo {pages} {113} (\bibinfo {year} {2012})},\ \Eprint {http://arxiv.org/abs/1208.6583} {arXiv:1208.6583 [hep-ph]} \BibitemShut {NoStop}%
\bibitem [{\citenamefont {Kurz}\ \emph {et~al.}(2014)\citenamefont {Kurz}, \citenamefont {Liu}, \citenamefont {Marquard},\ and\ \citenamefont {Steinhauser}}]{Kurz:2014wya}%
  \BibitemOpen
  \bibfield  {author} {\bibinfo {author} {\bibfnamefont {A.}~\bibnamefont {Kurz}}, \bibinfo {author} {\bibfnamefont {T.}~\bibnamefont {Liu}}, \bibinfo {author} {\bibfnamefont {P.}~\bibnamefont {Marquard}}, \ and\ \bibinfo {author} {\bibfnamefont {M.}~\bibnamefont {Steinhauser}},\ }\href {\doibase 10.1016/j.physletb.2014.05.043} {\bibfield  {journal} {\bibinfo  {journal} {Phys. Lett. B}\ }\textbf {\bibinfo {volume} {734}},\ \bibinfo {pages} {144} (\bibinfo {year} {2014})},\ \Eprint {http://arxiv.org/abs/1403.6400} {arXiv:1403.6400 [hep-ph]} \BibitemShut {NoStop}%
\bibitem [{\citenamefont {Kurz}\ \emph {et~al.}(2015)\citenamefont {Kurz}, \citenamefont {Liu}, \citenamefont {Marquard}, \citenamefont {Smirnov}, \citenamefont {Smirnov},\ and\ \citenamefont {Steinhauser}}]{Kurz:2015bia}%
  \BibitemOpen
  \bibfield  {author} {\bibinfo {author} {\bibfnamefont {A.}~\bibnamefont {Kurz}}, \bibinfo {author} {\bibfnamefont {T.}~\bibnamefont {Liu}}, \bibinfo {author} {\bibfnamefont {P.}~\bibnamefont {Marquard}}, \bibinfo {author} {\bibfnamefont {A.~V.}\ \bibnamefont {Smirnov}}, \bibinfo {author} {\bibfnamefont {V.~A.}\ \bibnamefont {Smirnov}}, \ and\ \bibinfo {author} {\bibfnamefont {M.}~\bibnamefont {Steinhauser}},\ }\href {\doibase 10.1103/PhysRevD.92.073019} {\bibfield  {journal} {\bibinfo  {journal} {Phys. Rev. D}\ }\textbf {\bibinfo {volume} {92}},\ \bibinfo {pages} {073019} (\bibinfo {year} {2015})},\ \Eprint {http://arxiv.org/abs/1508.00901} {arXiv:1508.00901 [hep-ph]} \BibitemShut {NoStop}%
\bibitem [{\citenamefont {Kurz}\ \emph {et~al.}(2016)\citenamefont {Kurz}, \citenamefont {Liu}, \citenamefont {Marquard}, \citenamefont {Smirnov}, \citenamefont {Smirnov},\ and\ \citenamefont {Steinhauser}}]{Kurz:2016bau}%
  \BibitemOpen
  \bibfield  {author} {\bibinfo {author} {\bibfnamefont {A.}~\bibnamefont {Kurz}}, \bibinfo {author} {\bibfnamefont {T.}~\bibnamefont {Liu}}, \bibinfo {author} {\bibfnamefont {P.}~\bibnamefont {Marquard}}, \bibinfo {author} {\bibfnamefont {A.}~\bibnamefont {Smirnov}}, \bibinfo {author} {\bibfnamefont {V.}~\bibnamefont {Smirnov}}, \ and\ \bibinfo {author} {\bibfnamefont {M.}~\bibnamefont {Steinhauser}},\ }\href {\doibase 10.1103/PhysRevD.93.053017} {\bibfield  {journal} {\bibinfo  {journal} {Phys. Rev. D}\ }\textbf {\bibinfo {volume} {93}},\ \bibinfo {pages} {053017} (\bibinfo {year} {2016})},\ \Eprint {http://arxiv.org/abs/1602.02785} {arXiv:1602.02785 [hep-ph]} \BibitemShut {NoStop}%
\bibitem [{\citenamefont {Liu}\ \emph {et~al.}(2019)\citenamefont {Liu}, \citenamefont {Wagner},\ and\ \citenamefont {Wang}}]{Liu:2018xkx}%
  \BibitemOpen
  \bibfield  {author} {\bibinfo {author} {\bibfnamefont {J.}~\bibnamefont {Liu}}, \bibinfo {author} {\bibfnamefont {C.~E.~M.}\ \bibnamefont {Wagner}}, \ and\ \bibinfo {author} {\bibfnamefont {X.-P.}\ \bibnamefont {Wang}},\ }\href {\doibase 10.1007/JHEP03(2019)008} {\bibfield  {journal} {\bibinfo  {journal} {JHEP}\ }\textbf {\bibinfo {volume} {03}},\ \bibinfo {pages} {008} (\bibinfo {year} {2019})},\ \Eprint {http://arxiv.org/abs/1810.11028} {arXiv:1810.11028 [hep-ph]} \BibitemShut {NoStop}%
\bibitem [{\citenamefont {Liu}\ \emph {et~al.}(2020)\citenamefont {Liu}, \citenamefont {McGinnis}, \citenamefont {Wagner},\ and\ \citenamefont {Wang}}]{Liu:2020qgx}%
  \BibitemOpen
  \bibfield  {author} {\bibinfo {author} {\bibfnamefont {J.}~\bibnamefont {Liu}}, \bibinfo {author} {\bibfnamefont {N.}~\bibnamefont {McGinnis}}, \bibinfo {author} {\bibfnamefont {C.~E.~M.}\ \bibnamefont {Wagner}}, \ and\ \bibinfo {author} {\bibfnamefont {X.-P.}\ \bibnamefont {Wang}},\ }\href {\doibase 10.1007/JHEP04(2020)197} {\bibfield  {journal} {\bibinfo  {journal} {JHEP}\ }\textbf {\bibinfo {volume} {04}},\ \bibinfo {pages} {197} (\bibinfo {year} {2020})},\ \Eprint {http://arxiv.org/abs/2001.06522} {arXiv:2001.06522 [hep-ph]} \BibitemShut {NoStop}%
\bibitem [{\citenamefont {Aebischer}\ \emph {et~al.}(2021)\citenamefont {Aebischer}, \citenamefont {Dekens}, \citenamefont {Jenkins}, \citenamefont {Manohar}, \citenamefont {Sengupta},\ and\ \citenamefont {Stoffer}}]{Aebischer:2021uvt}%
  \BibitemOpen
  \bibfield  {author} {\bibinfo {author} {\bibfnamefont {J.}~\bibnamefont {Aebischer}}, \bibinfo {author} {\bibfnamefont {W.}~\bibnamefont {Dekens}}, \bibinfo {author} {\bibfnamefont {E.~E.}\ \bibnamefont {Jenkins}}, \bibinfo {author} {\bibfnamefont {A.~V.}\ \bibnamefont {Manohar}}, \bibinfo {author} {\bibfnamefont {D.}~\bibnamefont {Sengupta}}, \ and\ \bibinfo {author} {\bibfnamefont {P.}~\bibnamefont {Stoffer}},\ }\href {\doibase 10.1007/JHEP07(2021)107} {\bibfield  {journal} {\bibinfo  {journal} {JHEP}\ }\textbf {\bibinfo {volume} {07}},\ \bibinfo {pages} {107} (\bibinfo {year} {2021})},\ \Eprint {http://arxiv.org/abs/2102.08954} {arXiv:2102.08954 [hep-ph]} \BibitemShut {NoStop}%
\bibitem [{\citenamefont {Li}\ \emph {et~al.}(2022)\citenamefont {Li}, \citenamefont {Xiao},\ and\ \citenamefont {Yang}}]{Li:2021koa}%
  \BibitemOpen
  \bibfield  {author} {\bibinfo {author} {\bibfnamefont {S.}~\bibnamefont {Li}}, \bibinfo {author} {\bibfnamefont {Y.}~\bibnamefont {Xiao}}, \ and\ \bibinfo {author} {\bibfnamefont {J.~M.}\ \bibnamefont {Yang}},\ }\href {\doibase 10.1140/epjc/s10052-022-10242-y} {\bibfield  {journal} {\bibinfo  {journal} {Eur. Phys. J. C}\ }\textbf {\bibinfo {volume} {82}},\ \bibinfo {pages} {276} (\bibinfo {year} {2022})},\ \Eprint {http://arxiv.org/abs/2107.04962} {arXiv:2107.04962 [hep-ph]} \BibitemShut {NoStop}%
\bibitem [{\citenamefont {Cirigliano}\ \emph {et~al.}(2021)\citenamefont {Cirigliano}, \citenamefont {Dekens}, \citenamefont {de~Vries}, \citenamefont {Fuyuto}, \citenamefont {Mereghetti},\ and\ \citenamefont {Ruiz}}]{Cirigliano:2021peb}%
  \BibitemOpen
  \bibfield  {author} {\bibinfo {author} {\bibfnamefont {V.}~\bibnamefont {Cirigliano}}, \bibinfo {author} {\bibfnamefont {W.}~\bibnamefont {Dekens}}, \bibinfo {author} {\bibfnamefont {J.}~\bibnamefont {de~Vries}}, \bibinfo {author} {\bibfnamefont {K.}~\bibnamefont {Fuyuto}}, \bibinfo {author} {\bibfnamefont {E.}~\bibnamefont {Mereghetti}}, \ and\ \bibinfo {author} {\bibfnamefont {R.}~\bibnamefont {Ruiz}},\ }\href {\doibase 10.1007/JHEP08(2021)103} {\bibfield  {journal} {\bibinfo  {journal} {JHEP}\ }\textbf {\bibinfo {volume} {08}},\ \bibinfo {pages} {103} (\bibinfo {year} {2021})},\ \Eprint {http://arxiv.org/abs/2105.11462} {arXiv:2105.11462 [hep-ph]} \BibitemShut {NoStop}%
\bibitem [{\citenamefont {Bhupal~Dev}\ \emph {et~al.}(2022)\citenamefont {Bhupal~Dev}, \citenamefont {Soni},\ and\ \citenamefont {Xu}}]{BhupalDev:2021ipu}%
  \BibitemOpen
  \bibfield  {author} {\bibinfo {author} {\bibfnamefont {P.~S.}\ \bibnamefont {Bhupal~Dev}}, \bibinfo {author} {\bibfnamefont {A.}~\bibnamefont {Soni}}, \ and\ \bibinfo {author} {\bibfnamefont {F.}~\bibnamefont {Xu}},\ }\href {\doibase 10.1103/PhysRevD.106.015014} {\bibfield  {journal} {\bibinfo  {journal} {Phys. Rev. D}\ }\textbf {\bibinfo {volume} {106}},\ \bibinfo {pages} {015014} (\bibinfo {year} {2022})},\ \Eprint {http://arxiv.org/abs/2106.15647} {arXiv:2106.15647 [hep-ph]} \BibitemShut {NoStop}%
\bibitem [{\citenamefont {Afik}\ \emph {et~al.}(2023)\citenamefont {Afik}, \citenamefont {Dev}, \citenamefont {Soni},\ and\ \citenamefont {Xu}}]{Afik:2022vpm}%
  \BibitemOpen
  \bibfield  {author} {\bibinfo {author} {\bibfnamefont {Y.}~\bibnamefont {Afik}}, \bibinfo {author} {\bibfnamefont {P.~S.~B.}\ \bibnamefont {Dev}}, \bibinfo {author} {\bibfnamefont {A.}~\bibnamefont {Soni}}, \ and\ \bibinfo {author} {\bibfnamefont {F.}~\bibnamefont {Xu}},\ }\href {\doibase 10.1016/j.physletb.2023.138032} {\bibfield  {journal} {\bibinfo  {journal} {Phys. Lett. B}\ }\textbf {\bibinfo {volume} {843}},\ \bibinfo {pages} {138032} (\bibinfo {year} {2023})},\ \Eprint {http://arxiv.org/abs/2212.06160} {arXiv:2212.06160 [hep-ph]} \BibitemShut {NoStop}%
\bibitem [{\citenamefont {Xu}(2023)}]{Xu:2023ene}%
  \BibitemOpen
  \bibfield  {author} {\bibinfo {author} {\bibfnamefont {F.}~\bibnamefont {Xu}},\ }\href {\doibase 10.1103/PhysRevD.108.036002} {\bibfield  {journal} {\bibinfo  {journal} {Phys. Rev. D}\ }\textbf {\bibinfo {volume} {108}},\ \bibinfo {pages} {036002} (\bibinfo {year} {2023})},\ \Eprint {http://arxiv.org/abs/2302.08653} {arXiv:2302.08653 [hep-ph]} \BibitemShut {NoStop}%
\bibitem [{\citenamefont {Wen}\ \emph {et~al.}(2023)\citenamefont {Wen}, \citenamefont {Yan}, \citenamefont {Yu},\ and\ \citenamefont {Yuan}}]{Wen:2023xxc}%
  \BibitemOpen
  \bibfield  {author} {\bibinfo {author} {\bibfnamefont {X.-K.}\ \bibnamefont {Wen}}, \bibinfo {author} {\bibfnamefont {B.}~\bibnamefont {Yan}}, \bibinfo {author} {\bibfnamefont {Z.}~\bibnamefont {Yu}}, \ and\ \bibinfo {author} {\bibfnamefont {C.~P.}\ \bibnamefont {Yuan}},\ }\href@noop {} {\  (\bibinfo {year} {2023})},\ \Eprint {http://arxiv.org/abs/2307.05236} {arXiv:2307.05236 [hep-ph]} \BibitemShut {NoStop}%
\bibitem [{\citenamefont {Cao}\ \emph {et~al.}(2023)\citenamefont {Cao}, \citenamefont {Meng},\ and\ \citenamefont {Yue}}]{Cao:2023juc}%
  \BibitemOpen
  \bibfield  {author} {\bibinfo {author} {\bibfnamefont {J.}~\bibnamefont {Cao}}, \bibinfo {author} {\bibfnamefont {L.}~\bibnamefont {Meng}}, \ and\ \bibinfo {author} {\bibfnamefont {Y.}~\bibnamefont {Yue}},\ }\href {\doibase 10.1103/PhysRevD.108.035043} {\bibfield  {journal} {\bibinfo  {journal} {Phys. Rev. D}\ }\textbf {\bibinfo {volume} {108}},\ \bibinfo {pages} {035043} (\bibinfo {year} {2023})},\ \Eprint {http://arxiv.org/abs/2306.06854} {arXiv:2306.06854 [hep-ph]} \BibitemShut {NoStop}%
\bibitem [{\citenamefont {Navas}\ \emph {et~al.}(2024)\citenamefont {Navas} \emph {et~al.}}]{ParticleDataGroup:2024cfk}%
  \BibitemOpen
  \bibfield  {author} {\bibinfo {author} {\bibfnamefont {S.}~\bibnamefont {Navas}} \emph {et~al.} (\bibinfo {collaboration} {Particle Data Group}),\ }\href {\doibase 10.1103/PhysRevD.110.030001} {\bibfield  {journal} {\bibinfo  {journal} {Phys. Rev. D}\ }\textbf {\bibinfo {volume} {110}},\ \bibinfo {pages} {030001} (\bibinfo {year} {2024})}\BibitemShut {NoStop}%
\bibitem [{\citenamefont {Aguillard}\ \emph {et~al.}(2023)\citenamefont {Aguillard} \emph {et~al.}}]{Muong-2:2023cdq}%
  \BibitemOpen
  \bibfield  {author} {\bibinfo {author} {\bibfnamefont {D.~P.}\ \bibnamefont {Aguillard}} \emph {et~al.} (\bibinfo {collaboration} {Muon g-2}),\ }\href@noop {} {\  (\bibinfo {year} {2023})},\ \Eprint {http://arxiv.org/abs/2308.06230} {arXiv:2308.06230 [hep-ex]} \BibitemShut {NoStop}%
\bibitem [{\citenamefont {Aguillard}\ \emph {et~al.}(2025)\citenamefont {Aguillard} \emph {et~al.}}]{Muong-2:2025xyk}%
  \BibitemOpen
  \bibfield  {author} {\bibinfo {author} {\bibfnamefont {D.~P.}\ \bibnamefont {Aguillard}} \emph {et~al.} (\bibinfo {collaboration} {Muon g-2}),\ }\href@noop {} {\  (\bibinfo {year} {2025})},\ \Eprint {http://arxiv.org/abs/2506.03069} {arXiv:2506.03069 [hep-ex]} \BibitemShut {NoStop}%
\bibitem [{\citenamefont {Aliberti}\ \emph {et~al.}(2025)\citenamefont {Aliberti} \emph {et~al.}}]{Aliberti:2025beg}%
  \BibitemOpen
  \bibfield  {author} {\bibinfo {author} {\bibfnamefont {R.}~\bibnamefont {Aliberti}} \emph {et~al.},\ }\href@noop {} {\  (\bibinfo {year} {2025})},\ \Eprint {http://arxiv.org/abs/2505.21476} {arXiv:2505.21476 [hep-ph]} \BibitemShut {NoStop}%
\bibitem [{\citenamefont {del Aguila}\ \emph {et~al.}(1991)\citenamefont {del Aguila}, \citenamefont {Cornet},\ and\ \citenamefont {Illana}}]{delAguila:1991rm}%
  \BibitemOpen
  \bibfield  {author} {\bibinfo {author} {\bibfnamefont {F.}~\bibnamefont {del Aguila}}, \bibinfo {author} {\bibfnamefont {F.}~\bibnamefont {Cornet}}, \ and\ \bibinfo {author} {\bibfnamefont {J.~I.}\ \bibnamefont {Illana}},\ }\href {\doibase 10.1016/0370-2693(91)91309-J} {\bibfield  {journal} {\bibinfo  {journal} {Phys. Lett. B}\ }\textbf {\bibinfo {volume} {271}},\ \bibinfo {pages} {256} (\bibinfo {year} {1991})}\BibitemShut {NoStop}%
\bibitem [{\citenamefont {Abdallah}\ \emph {et~al.}(2004)\citenamefont {Abdallah} \emph {et~al.}}]{DELPHI:2003nah}%
  \BibitemOpen
  \bibfield  {author} {\bibinfo {author} {\bibfnamefont {J.}~\bibnamefont {Abdallah}} \emph {et~al.} (\bibinfo {collaboration} {DELPHI}),\ }\href {\doibase 10.1140/epjc/s2004-01852-y} {\bibfield  {journal} {\bibinfo  {journal} {Eur. Phys. J. C}\ }\textbf {\bibinfo {volume} {35}},\ \bibinfo {pages} {159} (\bibinfo {year} {2004})},\ \Eprint {http://arxiv.org/abs/hep-ex/0406010} {arXiv:hep-ex/0406010} \BibitemShut {NoStop}%
\bibitem [{\citenamefont {Bernabeu}\ \emph {et~al.}(2008)\citenamefont {Bernabeu}, \citenamefont {Gonzalez-Sprinberg}, \citenamefont {Papavassiliou},\ and\ \citenamefont {Vidal}}]{Bernabeu:2007rr}%
  \BibitemOpen
  \bibfield  {author} {\bibinfo {author} {\bibfnamefont {J.}~\bibnamefont {Bernabeu}}, \bibinfo {author} {\bibfnamefont {G.~A.}\ \bibnamefont {Gonzalez-Sprinberg}}, \bibinfo {author} {\bibfnamefont {J.}~\bibnamefont {Papavassiliou}}, \ and\ \bibinfo {author} {\bibfnamefont {J.}~\bibnamefont {Vidal}},\ }\href {\doibase 10.1016/j.nuclphysb.2007.09.001} {\bibfield  {journal} {\bibinfo  {journal} {Nucl. Phys. B}\ }\textbf {\bibinfo {volume} {790}},\ \bibinfo {pages} {160} (\bibinfo {year} {2008})},\ \Eprint {http://arxiv.org/abs/0707.2496} {arXiv:0707.2496 [hep-ph]} \BibitemShut {NoStop}%
\bibitem [{\citenamefont {Atag}\ and\ \citenamefont {Billur}(2010)}]{Atag:2010ja}%
  \BibitemOpen
  \bibfield  {author} {\bibinfo {author} {\bibfnamefont {S.}~\bibnamefont {Atag}}\ and\ \bibinfo {author} {\bibfnamefont {A.~A.}\ \bibnamefont {Billur}},\ }\href {\doibase 10.1007/JHEP11(2010)060} {\bibfield  {journal} {\bibinfo  {journal} {JHEP}\ }\textbf {\bibinfo {volume} {11}},\ \bibinfo {pages} {060} (\bibinfo {year} {2010})},\ \Eprint {http://arxiv.org/abs/1005.2841} {arXiv:1005.2841 [hep-ph]} \BibitemShut {NoStop}%
\bibitem [{\citenamefont {Billur}\ and\ \citenamefont {Koksal}(2014)}]{Billur:2013rva}%
  \BibitemOpen
  \bibfield  {author} {\bibinfo {author} {\bibfnamefont {A.~A.}\ \bibnamefont {Billur}}\ and\ \bibinfo {author} {\bibfnamefont {M.}~\bibnamefont {Koksal}},\ }\href {\doibase 10.1103/PhysRevD.89.037301} {\bibfield  {journal} {\bibinfo  {journal} {Phys. Rev. D}\ }\textbf {\bibinfo {volume} {89}},\ \bibinfo {pages} {037301} (\bibinfo {year} {2014})},\ \Eprint {http://arxiv.org/abs/1306.5620} {arXiv:1306.5620 [hep-ph]} \BibitemShut {NoStop}%
\bibitem [{\citenamefont {Eidelman}\ \emph {et~al.}(2016)\citenamefont {Eidelman}, \citenamefont {Epifanov}, \citenamefont {Fael}, \citenamefont {Mercolli},\ and\ \citenamefont {Passera}}]{Eidelman:2016aih}%
  \BibitemOpen
  \bibfield  {author} {\bibinfo {author} {\bibfnamefont {S.}~\bibnamefont {Eidelman}}, \bibinfo {author} {\bibfnamefont {D.}~\bibnamefont {Epifanov}}, \bibinfo {author} {\bibfnamefont {M.}~\bibnamefont {Fael}}, \bibinfo {author} {\bibfnamefont {L.}~\bibnamefont {Mercolli}}, \ and\ \bibinfo {author} {\bibfnamefont {M.}~\bibnamefont {Passera}},\ }\href {\doibase 10.1007/JHEP03(2016)140} {\bibfield  {journal} {\bibinfo  {journal} {JHEP}\ }\textbf {\bibinfo {volume} {03}},\ \bibinfo {pages} {140} (\bibinfo {year} {2016})},\ \Eprint {http://arxiv.org/abs/1601.07987} {arXiv:1601.07987 [hep-ph]} \BibitemShut {NoStop}%
\bibitem [{\citenamefont {Chen}\ and\ \citenamefont {Wu}(2019)}]{Chen:2018cxt}%
  \BibitemOpen
  \bibfield  {author} {\bibinfo {author} {\bibfnamefont {X.}~\bibnamefont {Chen}}\ and\ \bibinfo {author} {\bibfnamefont {Y.}~\bibnamefont {Wu}},\ }\href {\doibase 10.1007/JHEP10(2019)089} {\bibfield  {journal} {\bibinfo  {journal} {JHEP}\ }\textbf {\bibinfo {volume} {10}},\ \bibinfo {pages} {089} (\bibinfo {year} {2019})},\ \Eprint {http://arxiv.org/abs/1803.00501} {arXiv:1803.00501 [hep-ph]} \BibitemShut {NoStop}%
\bibitem [{\citenamefont {Fu}\ \emph {et~al.}(2019)\citenamefont {Fu}, \citenamefont {Giorgi}, \citenamefont {Henry}, \citenamefont {Marangotto}, \citenamefont {Vidal}, \citenamefont {Merli}, \citenamefont {Neri},\ and\ \citenamefont {Ruiz~Vidal}}]{Fu:2019utm}%
  \BibitemOpen
  \bibfield  {author} {\bibinfo {author} {\bibfnamefont {J.}~\bibnamefont {Fu}}, \bibinfo {author} {\bibfnamefont {M.~A.}\ \bibnamefont {Giorgi}}, \bibinfo {author} {\bibfnamefont {L.}~\bibnamefont {Henry}}, \bibinfo {author} {\bibfnamefont {D.}~\bibnamefont {Marangotto}}, \bibinfo {author} {\bibfnamefont {F.~M.}\ \bibnamefont {Vidal}}, \bibinfo {author} {\bibfnamefont {A.}~\bibnamefont {Merli}}, \bibinfo {author} {\bibfnamefont {N.}~\bibnamefont {Neri}}, \ and\ \bibinfo {author} {\bibfnamefont {J.}~\bibnamefont {Ruiz~Vidal}},\ }\href {\doibase 10.1103/PhysRevLett.123.011801} {\bibfield  {journal} {\bibinfo  {journal} {Phys. Rev. Lett.}\ }\textbf {\bibinfo {volume} {123}},\ \bibinfo {pages} {011801} (\bibinfo {year} {2019})},\ \Eprint {http://arxiv.org/abs/1901.04003} {arXiv:1901.04003 [hep-ex]} \BibitemShut {NoStop}%
\bibitem [{\citenamefont {Beresford}\ and\ \citenamefont {Liu}(2020)}]{Beresford:2019gww}%
  \BibitemOpen
  \bibfield  {author} {\bibinfo {author} {\bibfnamefont {L.}~\bibnamefont {Beresford}}\ and\ \bibinfo {author} {\bibfnamefont {J.}~\bibnamefont {Liu}},\ }\href {\doibase 10.1103/PhysRevD.102.113008} {\bibfield  {journal} {\bibinfo  {journal} {Phys. Rev. D}\ }\textbf {\bibinfo {volume} {102}},\ \bibinfo {pages} {113008} (\bibinfo {year} {2020})},\ \bibinfo {note} {[Erratum: Phys.Rev.D 106, 039902 (2022)]},\ \Eprint {http://arxiv.org/abs/1908.05180} {arXiv:1908.05180 [hep-ph]} \BibitemShut {NoStop}%
\bibitem [{\citenamefont {Dyndal}\ \emph {et~al.}(2020)\citenamefont {Dyndal}, \citenamefont {Klusek-Gawenda}, \citenamefont {Schott},\ and\ \citenamefont {Szczurek}}]{Dyndal:2020yen}%
  \BibitemOpen
  \bibfield  {author} {\bibinfo {author} {\bibfnamefont {M.}~\bibnamefont {Dyndal}}, \bibinfo {author} {\bibfnamefont {M.}~\bibnamefont {Klusek-Gawenda}}, \bibinfo {author} {\bibfnamefont {M.}~\bibnamefont {Schott}}, \ and\ \bibinfo {author} {\bibfnamefont {A.}~\bibnamefont {Szczurek}},\ }\href {\doibase 10.1016/j.physletb.2020.135682} {\bibfield  {journal} {\bibinfo  {journal} {Phys. Lett. B}\ }\textbf {\bibinfo {volume} {809}},\ \bibinfo {pages} {135682} (\bibinfo {year} {2020})},\ \Eprint {http://arxiv.org/abs/2002.05503} {arXiv:2002.05503 [hep-ph]} \BibitemShut {NoStop}%
\bibitem [{\citenamefont {Inami}\ \emph {et~al.}(2022)\citenamefont {Inami} \emph {et~al.}}]{Belle:2021ybo}%
  \BibitemOpen
  \bibfield  {author} {\bibinfo {author} {\bibfnamefont {K.}~\bibnamefont {Inami}} \emph {et~al.} (\bibinfo {collaboration} {Belle}),\ }\href {\doibase 10.1007/JHEP04(2022)110} {\bibfield  {journal} {\bibinfo  {journal} {JHEP}\ }\textbf {\bibinfo {volume} {04}},\ \bibinfo {pages} {110} (\bibinfo {year} {2022})},\ \Eprint {http://arxiv.org/abs/2108.11543} {arXiv:2108.11543 [hep-ex]} \BibitemShut {NoStop}%
\bibitem [{\citenamefont {Crivellin}\ \emph {et~al.}(2022)\citenamefont {Crivellin}, \citenamefont {Hoferichter},\ and\ \citenamefont {Roney}}]{Crivellin:2021spu}%
  \BibitemOpen
  \bibfield  {author} {\bibinfo {author} {\bibfnamefont {A.}~\bibnamefont {Crivellin}}, \bibinfo {author} {\bibfnamefont {M.}~\bibnamefont {Hoferichter}}, \ and\ \bibinfo {author} {\bibfnamefont {J.~M.}\ \bibnamefont {Roney}},\ }\href {\doibase 10.1103/PhysRevD.106.093007} {\bibfield  {journal} {\bibinfo  {journal} {Phys. Rev. D}\ }\textbf {\bibinfo {volume} {106}},\ \bibinfo {pages} {093007} (\bibinfo {year} {2022})},\ \Eprint {http://arxiv.org/abs/2111.10378} {arXiv:2111.10378 [hep-ph]} \BibitemShut {NoStop}%
\bibitem [{\citenamefont {Asner}\ \emph {et~al.}(2022)\citenamefont {Asner} \emph {et~al.}}]{USBelleIIGroup:2022qro}%
  \BibitemOpen
  \bibfield  {author} {\bibinfo {author} {\bibfnamefont {D.~M.}\ \bibnamefont {Asner}} \emph {et~al.} (\bibinfo {collaboration} {US Belle II Group, Belle II/SuperKEKB e- Polarization Upgrade Working Group}),\ }in\ \href@noop {} {\emph {\bibinfo {booktitle} {{Snowmass 2021}}}}\ (\bibinfo {year} {2022})\ \Eprint {http://arxiv.org/abs/2205.12847} {arXiv:2205.12847 [physics.acc-ph]} \BibitemShut {NoStop}%
\bibitem [{\citenamefont {Aad}\ \emph {et~al.}(2023)\citenamefont {Aad} \emph {et~al.}}]{ATLAS:2022ryk}%
  \BibitemOpen
  \bibfield  {author} {\bibinfo {author} {\bibfnamefont {G.}~\bibnamefont {Aad}} \emph {et~al.} (\bibinfo {collaboration} {ATLAS}),\ }\href {\doibase 10.1103/PhysRevLett.131.151802} {\bibfield  {journal} {\bibinfo  {journal} {Phys. Rev. Lett.}\ }\textbf {\bibinfo {volume} {131}},\ \bibinfo {pages} {151802} (\bibinfo {year} {2023})},\ \Eprint {http://arxiv.org/abs/2204.13478} {arXiv:2204.13478 [hep-ex]} \BibitemShut {NoStop}%
\bibitem [{\citenamefont {Tumasyan}\ \emph {et~al.}(2023)\citenamefont {Tumasyan} \emph {et~al.}}]{CMS:2022arf}%
  \BibitemOpen
  \bibfield  {author} {\bibinfo {author} {\bibfnamefont {A.}~\bibnamefont {Tumasyan}} \emph {et~al.} (\bibinfo {collaboration} {CMS}),\ }\href {\doibase 10.1103/PhysRevLett.131.151803} {\bibfield  {journal} {\bibinfo  {journal} {Phys. Rev. Lett.}\ }\textbf {\bibinfo {volume} {131}},\ \bibinfo {pages} {151803} (\bibinfo {year} {2023})},\ \Eprint {http://arxiv.org/abs/2206.05192} {arXiv:2206.05192 [nucl-ex]} \BibitemShut {NoStop}%
\bibitem [{\citenamefont {Verducci}\ \emph {et~al.}(2024)\citenamefont {Verducci}, \citenamefont {Roda}, \citenamefont {Cavasinni},\ and\ \citenamefont {Vignaroli}}]{Verducci:2023cgx}%
  \BibitemOpen
  \bibfield  {author} {\bibinfo {author} {\bibfnamefont {M.}~\bibnamefont {Verducci}}, \bibinfo {author} {\bibfnamefont {C.}~\bibnamefont {Roda}}, \bibinfo {author} {\bibfnamefont {V.}~\bibnamefont {Cavasinni}}, \ and\ \bibinfo {author} {\bibfnamefont {N.}~\bibnamefont {Vignaroli}},\ }\href {\doibase 10.1103/PhysRevD.110.052001} {\bibfield  {journal} {\bibinfo  {journal} {Phys. Rev. D}\ }\textbf {\bibinfo {volume} {110}},\ \bibinfo {pages} {052001} (\bibinfo {year} {2024})},\ \Eprint {http://arxiv.org/abs/2307.15160} {arXiv:2307.15160 [hep-ph]} \BibitemShut {NoStop}%
\bibitem [{\citenamefont {Denizli}\ \emph {et~al.}(2025)\citenamefont {Denizli}, \citenamefont {Senol},\ and\ \citenamefont {K\"oksal}}]{Denizli:2024uwv}%
  \BibitemOpen
  \bibfield  {author} {\bibinfo {author} {\bibfnamefont {H.}~\bibnamefont {Denizli}}, \bibinfo {author} {\bibfnamefont {A.}~\bibnamefont {Senol}}, \ and\ \bibinfo {author} {\bibfnamefont {M.}~\bibnamefont {K\"oksal}},\ }\href {\doibase 10.1016/j.cjph.2025.04.020} {\bibfield  {journal} {\bibinfo  {journal} {Chin. J. Phys.}\ }\textbf {\bibinfo {volume} {95}},\ \bibinfo {pages} {1250} (\bibinfo {year} {2025})},\ \Eprint {http://arxiv.org/abs/2408.16106} {arXiv:2408.16106 [hep-ph]} \BibitemShut {NoStop}%
\bibitem [{\citenamefont {Hayrapetyan}\ \emph {et~al.}(2024)\citenamefont {Hayrapetyan} \emph {et~al.}}]{CMS:2024qjo}%
  \BibitemOpen
  \bibfield  {author} {\bibinfo {author} {\bibfnamefont {A.}~\bibnamefont {Hayrapetyan}} \emph {et~al.} (\bibinfo {collaboration} {CMS}),\ }\href {\doibase 10.1088/1361-6633/ad6fcb} {\bibfield  {journal} {\bibinfo  {journal} {Rept. Prog. Phys.}\ }\textbf {\bibinfo {volume} {87}},\ \bibinfo {pages} {107801} (\bibinfo {year} {2024})},\ \Eprint {http://arxiv.org/abs/2406.03975} {arXiv:2406.03975 [hep-ex]} \BibitemShut {NoStop}%
\bibitem [{\citenamefont {Gogniat}\ \emph {et~al.}(2025)\citenamefont {Gogniat}, \citenamefont {Hoferichter},\ and\ \citenamefont {Ulrich}}]{Gogniat:2025eom}%
  \BibitemOpen
  \bibfield  {author} {\bibinfo {author} {\bibfnamefont {J.}~\bibnamefont {Gogniat}}, \bibinfo {author} {\bibfnamefont {M.}~\bibnamefont {Hoferichter}}, \ and\ \bibinfo {author} {\bibfnamefont {Y.}~\bibnamefont {Ulrich}},\ }\href@noop {} {\  (\bibinfo {year} {2025})},\ \Eprint {http://arxiv.org/abs/2505.09678} {arXiv:2505.09678 [hep-ph]} \BibitemShut {NoStop}%
\bibitem [{\citenamefont {Baltz}(2008)}]{Baltz:2007kq}%
  \BibitemOpen
  \bibfield  {author} {\bibinfo {author} {\bibfnamefont {A.~J.}\ \bibnamefont {Baltz}},\ }\href {\doibase 10.1016/j.physrep.2007.12.001} {\bibfield  {journal} {\bibinfo  {journal} {Phys. Rept.}\ }\textbf {\bibinfo {volume} {458}},\ \bibinfo {pages} {1} (\bibinfo {year} {2008})},\ \Eprint {http://arxiv.org/abs/0706.3356} {arXiv:0706.3356 [nucl-ex]} \BibitemShut {NoStop}%
\bibitem [{\citenamefont {Li}\ \emph {et~al.}(2019)\citenamefont {Li}, \citenamefont {Zhou},\ and\ \citenamefont {Zhou}}]{Li:2019yzy}%
  \BibitemOpen
  \bibfield  {author} {\bibinfo {author} {\bibfnamefont {C.}~\bibnamefont {Li}}, \bibinfo {author} {\bibfnamefont {J.}~\bibnamefont {Zhou}}, \ and\ \bibinfo {author} {\bibfnamefont {Y.-J.}\ \bibnamefont {Zhou}},\ }\href {\doibase 10.1016/j.physletb.2019.07.005} {\bibfield  {journal} {\bibinfo  {journal} {Phys. Lett. B}\ }\textbf {\bibinfo {volume} {795}},\ \bibinfo {pages} {576} (\bibinfo {year} {2019})},\ \Eprint {http://arxiv.org/abs/1903.10084} {arXiv:1903.10084 [hep-ph]} \BibitemShut {NoStop}%
\bibitem [{\citenamefont {Shao}\ \emph {et~al.}(2023{\natexlab{a}})\citenamefont {Shao}, \citenamefont {Yan}, \citenamefont {Yuan},\ and\ \citenamefont {Zhang}}]{Shao:2023bga}%
  \BibitemOpen
  \bibfield  {author} {\bibinfo {author} {\bibfnamefont {D.~Y.}\ \bibnamefont {Shao}}, \bibinfo {author} {\bibfnamefont {B.}~\bibnamefont {Yan}}, \bibinfo {author} {\bibfnamefont {S.-R.}\ \bibnamefont {Yuan}}, \ and\ \bibinfo {author} {\bibfnamefont {C.}~\bibnamefont {Zhang}},\ }\href@noop {} {\  (\bibinfo {year} {2023}{\natexlab{a}})},\ \Eprint {http://arxiv.org/abs/2310.14153} {arXiv:2310.14153 [hep-ph]} \BibitemShut {NoStop}%
\bibitem [{\citenamefont {Bertulani}\ and\ \citenamefont {Baur}(1988)}]{Bertulani:1987tz}%
  \BibitemOpen
  \bibfield  {author} {\bibinfo {author} {\bibfnamefont {C.~A.}\ \bibnamefont {Bertulani}}\ and\ \bibinfo {author} {\bibfnamefont {G.}~\bibnamefont {Baur}},\ }\href {\doibase 10.1016/0370-1573(88)90142-1} {\bibfield  {journal} {\bibinfo  {journal} {Phys. Rept.}\ }\textbf {\bibinfo {volume} {163}},\ \bibinfo {pages} {299} (\bibinfo {year} {1988})}\BibitemShut {NoStop}%
\bibitem [{\citenamefont {Vidovic}\ \emph {et~al.}(1993)\citenamefont {Vidovic}, \citenamefont {Greiner}, \citenamefont {Best},\ and\ \citenamefont {Soff}}]{Vidovic:1992ik}%
  \BibitemOpen
  \bibfield  {author} {\bibinfo {author} {\bibfnamefont {M.}~\bibnamefont {Vidovic}}, \bibinfo {author} {\bibfnamefont {M.}~\bibnamefont {Greiner}}, \bibinfo {author} {\bibfnamefont {C.}~\bibnamefont {Best}}, \ and\ \bibinfo {author} {\bibfnamefont {G.}~\bibnamefont {Soff}},\ }\href {\doibase 10.1103/PhysRevC.47.2308} {\bibfield  {journal} {\bibinfo  {journal} {Phys. Rev. C}\ }\textbf {\bibinfo {volume} {47}},\ \bibinfo {pages} {2308} (\bibinfo {year} {1993})}\BibitemShut {NoStop}%
\bibitem [{\citenamefont {Wang}\ \emph {et~al.}(2023)\citenamefont {Wang}, \citenamefont {Brandenburg}, \citenamefont {Ruan}, \citenamefont {Shao}, \citenamefont {Xu}, \citenamefont {Yang},\ and\ \citenamefont {Zha}}]{Wang:2022ihj}%
  \BibitemOpen
  \bibfield  {author} {\bibinfo {author} {\bibfnamefont {X.}~\bibnamefont {Wang}}, \bibinfo {author} {\bibfnamefont {J.~D.}\ \bibnamefont {Brandenburg}}, \bibinfo {author} {\bibfnamefont {L.}~\bibnamefont {Ruan}}, \bibinfo {author} {\bibfnamefont {F.}~\bibnamefont {Shao}}, \bibinfo {author} {\bibfnamefont {Z.}~\bibnamefont {Xu}}, \bibinfo {author} {\bibfnamefont {C.}~\bibnamefont {Yang}}, \ and\ \bibinfo {author} {\bibfnamefont {W.}~\bibnamefont {Zha}},\ }\href {\doibase 10.1103/PhysRevC.107.044906} {\bibfield  {journal} {\bibinfo  {journal} {Phys. Rev. C}\ }\textbf {\bibinfo {volume} {107}},\ \bibinfo {pages} {044906} (\bibinfo {year} {2023})},\ \Eprint {http://arxiv.org/abs/2207.05595} {arXiv:2207.05595 [nucl-th]} \BibitemShut {NoStop}%
\bibitem [{\citenamefont {Achasov}\ \emph {et~al.}(2024)\citenamefont {Achasov} \emph {et~al.}}]{Achasov:2023gey}%
  \BibitemOpen
  \bibfield  {author} {\bibinfo {author} {\bibfnamefont {M.}~\bibnamefont {Achasov}} \emph {et~al.},\ }\href {\doibase 10.1007/s11467-023-1333-z} {\bibfield  {journal} {\bibinfo  {journal} {Front. Phys. (Beijing)}\ }\textbf {\bibinfo {volume} {19}},\ \bibinfo {pages} {14701} (\bibinfo {year} {2024})},\ \Eprint {http://arxiv.org/abs/2303.15790} {arXiv:2303.15790 [hep-ex]} \BibitemShut {NoStop}%
\bibitem [{\citenamefont {Barnyakov}(2020)}]{Barnyakov:2020vob}%
  \BibitemOpen
  \bibfield  {author} {\bibinfo {author} {\bibfnamefont {A.~Y.}\ \bibnamefont {Barnyakov}} (\bibinfo {collaboration} {Super Charm-Tau Factory}),\ }\href {\doibase 10.1088/1742-6596/1561/1/012004} {\bibfield  {journal} {\bibinfo  {journal} {J. Phys. Conf. Ser.}\ }\textbf {\bibinfo {volume} {1561}},\ \bibinfo {pages} {012004} (\bibinfo {year} {2020})}\BibitemShut {NoStop}%
\bibitem [{\citenamefont {Eidelman}\ and\ \citenamefont {Passera}(2007)}]{Eidelman:2007sb}%
  \BibitemOpen
  \bibfield  {author} {\bibinfo {author} {\bibfnamefont {S.}~\bibnamefont {Eidelman}}\ and\ \bibinfo {author} {\bibfnamefont {M.}~\bibnamefont {Passera}},\ }\href {\doibase 10.1142/S0217732307022694} {\bibfield  {journal} {\bibinfo  {journal} {Mod. Phys. Lett. A}\ }\textbf {\bibinfo {volume} {22}},\ \bibinfo {pages} {159} (\bibinfo {year} {2007})},\ \Eprint {http://arxiv.org/abs/hep-ph/0701260} {arXiv:hep-ph/0701260} \BibitemShut {NoStop}%
\bibitem [{\citenamefont {Schwinger}(1948)}]{Schwinger:1948iu}%
  \BibitemOpen
  \bibfield  {author} {\bibinfo {author} {\bibfnamefont {J.~S.}\ \bibnamefont {Schwinger}},\ }\href {\doibase 10.1103/PhysRev.73.416} {\bibfield  {journal} {\bibinfo  {journal} {Phys. Rev.}\ }\textbf {\bibinfo {volume} {73}},\ \bibinfo {pages} {416} (\bibinfo {year} {1948})}\BibitemShut {NoStop}%
\bibitem [{\citenamefont {Yamaguchi}\ and\ \citenamefont {Yamanaka}(2020)}]{Yamaguchi:2020eub}%
  \BibitemOpen
  \bibfield  {author} {\bibinfo {author} {\bibfnamefont {Y.}~\bibnamefont {Yamaguchi}}\ and\ \bibinfo {author} {\bibfnamefont {N.}~\bibnamefont {Yamanaka}},\ }\href {\doibase 10.1103/PhysRevLett.125.241802} {\bibfield  {journal} {\bibinfo  {journal} {Phys. Rev. Lett.}\ }\textbf {\bibinfo {volume} {125}},\ \bibinfo {pages} {241802} (\bibinfo {year} {2020})},\ \Eprint {http://arxiv.org/abs/2003.08195} {arXiv:2003.08195 [hep-ph]} \BibitemShut {NoStop}%
\bibitem [{\citenamefont {Yamaguchi}\ and\ \citenamefont {Yamanaka}(2021)}]{Yamaguchi:2020dsy}%
  \BibitemOpen
  \bibfield  {author} {\bibinfo {author} {\bibfnamefont {Y.}~\bibnamefont {Yamaguchi}}\ and\ \bibinfo {author} {\bibfnamefont {N.}~\bibnamefont {Yamanaka}},\ }\href {\doibase 10.1103/PhysRevD.103.013001} {\bibfield  {journal} {\bibinfo  {journal} {Phys. Rev. D}\ }\textbf {\bibinfo {volume} {103}},\ \bibinfo {pages} {013001} (\bibinfo {year} {2021})},\ \Eprint {http://arxiv.org/abs/2006.00281} {arXiv:2006.00281 [hep-ph]} \BibitemShut {NoStop}%
\bibitem [{\citenamefont {von Weizsacker}(1934)}]{vonWeizsacker:1934nji}%
  \BibitemOpen
  \bibfield  {author} {\bibinfo {author} {\bibfnamefont {C.~F.}\ \bibnamefont {von Weizsacker}},\ }\href {\doibase 10.1007/BF01333110} {\bibfield  {journal} {\bibinfo  {journal} {Z. Phys.}\ }\textbf {\bibinfo {volume} {88}},\ \bibinfo {pages} {612} (\bibinfo {year} {1934})}\BibitemShut {NoStop}%
\bibitem [{\citenamefont {Williams}(1933)}]{Williams:1933}%
  \BibitemOpen
  \bibfield  {author} {\bibinfo {author} {\bibfnamefont {E.~J.}\ \bibnamefont {Williams}},\ }\href {\doibase 10.1098/rspa.1933.0012} {\bibfield  {journal} {\bibinfo  {journal} {Proc. R. Soc. Lond.}\ }\textbf {\bibinfo {volume} {A139}},\ \bibinfo {pages} {163} (\bibinfo {year} {1933})}\BibitemShut {NoStop}%
\bibitem [{\citenamefont {Williams}(1934)}]{Williams:1934ad}%
  \BibitemOpen
  \bibfield  {author} {\bibinfo {author} {\bibfnamefont {E.~J.}\ \bibnamefont {Williams}},\ }\href {\doibase 10.1103/PhysRev.45.729} {\bibfield  {journal} {\bibinfo  {journal} {Phys. Rev.}\ }\textbf {\bibinfo {volume} {45}},\ \bibinfo {pages} {729} (\bibinfo {year} {1934})}\BibitemShut {NoStop}%
\bibitem [{\citenamefont {Krauss}\ \emph {et~al.}(1997)\citenamefont {Krauss}, \citenamefont {Greiner},\ and\ \citenamefont {Soff}}]{Krauss:1997vr}%
  \BibitemOpen
  \bibfield  {author} {\bibinfo {author} {\bibfnamefont {F.}~\bibnamefont {Krauss}}, \bibinfo {author} {\bibfnamefont {M.}~\bibnamefont {Greiner}}, \ and\ \bibinfo {author} {\bibfnamefont {G.}~\bibnamefont {Soff}},\ }\href {\doibase 10.1016/S0146-6410(97)00049-5} {\bibfield  {journal} {\bibinfo  {journal} {Prog. Part. Nucl. Phys.}\ }\textbf {\bibinfo {volume} {39}},\ \bibinfo {pages} {503} (\bibinfo {year} {1997})}\BibitemShut {NoStop}%
\bibitem [{\citenamefont {Adam}\ \emph {et~al.}(2021)\citenamefont {Adam} \emph {et~al.}}]{STAR:2019wlg}%
  \BibitemOpen
  \bibfield  {author} {\bibinfo {author} {\bibfnamefont {J.}~\bibnamefont {Adam}} \emph {et~al.} (\bibinfo {collaboration} {STAR}),\ }\href {\doibase 10.1103/PhysRevLett.127.052302} {\bibfield  {journal} {\bibinfo  {journal} {Phys. Rev. Lett.}\ }\textbf {\bibinfo {volume} {127}},\ \bibinfo {pages} {052302} (\bibinfo {year} {2021})},\ \Eprint {http://arxiv.org/abs/1910.12400} {arXiv:1910.12400 [nucl-ex]} \BibitemShut {NoStop}%
\bibitem [{\citenamefont {Li}\ \emph {et~al.}(2020)\citenamefont {Li}, \citenamefont {Zhou},\ and\ \citenamefont {Zhou}}]{Li:2019sin}%
  \BibitemOpen
  \bibfield  {author} {\bibinfo {author} {\bibfnamefont {C.}~\bibnamefont {Li}}, \bibinfo {author} {\bibfnamefont {J.}~\bibnamefont {Zhou}}, \ and\ \bibinfo {author} {\bibfnamefont {Y.-J.}\ \bibnamefont {Zhou}},\ }\href {\doibase 10.1103/PhysRevD.101.034015} {\bibfield  {journal} {\bibinfo  {journal} {Phys. Rev. D}\ }\textbf {\bibinfo {volume} {101}},\ \bibinfo {pages} {034015} (\bibinfo {year} {2020})},\ \Eprint {http://arxiv.org/abs/1911.00237} {arXiv:1911.00237 [hep-ph]} \BibitemShut {NoStop}%
\bibitem [{\citenamefont {Xiao}\ \emph {et~al.}(2020)\citenamefont {Xiao}, \citenamefont {Yuan},\ and\ \citenamefont {Zhou}}]{Xiao:2020ddm}%
  \BibitemOpen
  \bibfield  {author} {\bibinfo {author} {\bibfnamefont {B.-W.}\ \bibnamefont {Xiao}}, \bibinfo {author} {\bibfnamefont {F.}~\bibnamefont {Yuan}}, \ and\ \bibinfo {author} {\bibfnamefont {J.}~\bibnamefont {Zhou}},\ }\href {\doibase 10.1103/PhysRevLett.125.232301} {\bibfield  {journal} {\bibinfo  {journal} {Phys. Rev. Lett.}\ }\textbf {\bibinfo {volume} {125}},\ \bibinfo {pages} {232301} (\bibinfo {year} {2020})},\ \Eprint {http://arxiv.org/abs/2003.06352} {arXiv:2003.06352 [hep-ph]} \BibitemShut {NoStop}%
\bibitem [{\citenamefont {Klein}\ \emph {et~al.}(2020)\citenamefont {Klein}, \citenamefont {Mueller}, \citenamefont {Xiao},\ and\ \citenamefont {Yuan}}]{Klein:2020jom}%
  \BibitemOpen
  \bibfield  {author} {\bibinfo {author} {\bibfnamefont {S.}~\bibnamefont {Klein}}, \bibinfo {author} {\bibfnamefont {A.~H.}\ \bibnamefont {Mueller}}, \bibinfo {author} {\bibfnamefont {B.-W.}\ \bibnamefont {Xiao}}, \ and\ \bibinfo {author} {\bibfnamefont {F.}~\bibnamefont {Yuan}},\ }\href {\doibase 10.1103/PhysRevD.102.094013} {\bibfield  {journal} {\bibinfo  {journal} {Phys. Rev. D}\ }\textbf {\bibinfo {volume} {102}},\ \bibinfo {pages} {094013} (\bibinfo {year} {2020})},\ \Eprint {http://arxiv.org/abs/2003.02947} {arXiv:2003.02947 [hep-ph]} \BibitemShut {NoStop}%
\bibitem [{\citenamefont {Zhao}\ \emph {et~al.}(2024)\citenamefont {Zhao}, \citenamefont {Chen}, \citenamefont {Huang},\ and\ \citenamefont {Ma}}]{Zhao:2022dac}%
  \BibitemOpen
  \bibfield  {author} {\bibinfo {author} {\bibfnamefont {J.}~\bibnamefont {Zhao}}, \bibinfo {author} {\bibfnamefont {J.-H.}\ \bibnamefont {Chen}}, \bibinfo {author} {\bibfnamefont {X.-G.}\ \bibnamefont {Huang}}, \ and\ \bibinfo {author} {\bibfnamefont {Y.-G.}\ \bibnamefont {Ma}},\ }\href {\doibase 10.1007/s41365-024-01374-9} {\bibfield  {journal} {\bibinfo  {journal} {Nucl. Sci. Tech.}\ }\textbf {\bibinfo {volume} {35}},\ \bibinfo {pages} {20} (\bibinfo {year} {2024})},\ \Eprint {http://arxiv.org/abs/2211.03968} {arXiv:2211.03968 [nucl-th]} \BibitemShut {NoStop}%
\bibitem [{\citenamefont {Brandenburg}\ \emph {et~al.}(2022)\citenamefont {Brandenburg}, \citenamefont {Seger}, \citenamefont {Xu},\ and\ \citenamefont {Zha}}]{Brandenburg:2022tna}%
  \BibitemOpen
  \bibfield  {author} {\bibinfo {author} {\bibfnamefont {J.~D.}\ \bibnamefont {Brandenburg}}, \bibinfo {author} {\bibfnamefont {J.}~\bibnamefont {Seger}}, \bibinfo {author} {\bibfnamefont {Z.}~\bibnamefont {Xu}}, \ and\ \bibinfo {author} {\bibfnamefont {W.}~\bibnamefont {Zha}},\ }\href@noop {} {\  (\bibinfo {year} {2022})},\ \Eprint {http://arxiv.org/abs/2208.14943} {arXiv:2208.14943 [hep-ph]} \BibitemShut {NoStop}%
\bibitem [{\citenamefont {Zha}\ \emph {et~al.}(2020)\citenamefont {Zha}, \citenamefont {Brandenburg}, \citenamefont {Tang},\ and\ \citenamefont {Xu}}]{Zha:2018tlq}%
  \BibitemOpen
  \bibfield  {author} {\bibinfo {author} {\bibfnamefont {W.}~\bibnamefont {Zha}}, \bibinfo {author} {\bibfnamefont {J.~D.}\ \bibnamefont {Brandenburg}}, \bibinfo {author} {\bibfnamefont {Z.}~\bibnamefont {Tang}}, \ and\ \bibinfo {author} {\bibfnamefont {Z.}~\bibnamefont {Xu}},\ }\href {\doibase 10.1016/j.physletb.2019.135089} {\bibfield  {journal} {\bibinfo  {journal} {Phys. Lett. B}\ }\textbf {\bibinfo {volume} {800}},\ \bibinfo {pages} {135089} (\bibinfo {year} {2020})},\ \Eprint {http://arxiv.org/abs/1812.02820} {arXiv:1812.02820 [nucl-th]} \BibitemShut {NoStop}%
\bibitem [{\citenamefont {Wang}\ \emph {et~al.}(2021)\citenamefont {Wang}, \citenamefont {Pu},\ and\ \citenamefont {Wang}}]{Wang:2021kxm}%
  \BibitemOpen
  \bibfield  {author} {\bibinfo {author} {\bibfnamefont {R.-j.}\ \bibnamefont {Wang}}, \bibinfo {author} {\bibfnamefont {S.}~\bibnamefont {Pu}}, \ and\ \bibinfo {author} {\bibfnamefont {Q.}~\bibnamefont {Wang}},\ }\href {\doibase 10.1103/PhysRevD.104.056011} {\bibfield  {journal} {\bibinfo  {journal} {Phys. Rev. D}\ }\textbf {\bibinfo {volume} {104}},\ \bibinfo {pages} {056011} (\bibinfo {year} {2021})},\ \Eprint {http://arxiv.org/abs/2106.05462} {arXiv:2106.05462 [hep-ph]} \BibitemShut {NoStop}%
\bibitem [{\citenamefont {Wang}\ \emph {et~al.}(2022)\citenamefont {Wang}, \citenamefont {Lin}, \citenamefont {Pu}, \citenamefont {Zhang},\ and\ \citenamefont {Wang}}]{Wang:2022gkd}%
  \BibitemOpen
  \bibfield  {author} {\bibinfo {author} {\bibfnamefont {R.-j.}\ \bibnamefont {Wang}}, \bibinfo {author} {\bibfnamefont {S.}~\bibnamefont {Lin}}, \bibinfo {author} {\bibfnamefont {S.}~\bibnamefont {Pu}}, \bibinfo {author} {\bibfnamefont {Y.-f.}\ \bibnamefont {Zhang}}, \ and\ \bibinfo {author} {\bibfnamefont {Q.}~\bibnamefont {Wang}},\ }\href@noop {} {\  (\bibinfo {year} {2022})},\ \Eprint {http://arxiv.org/abs/2204.02761} {arXiv:2204.02761 [hep-ph]} \BibitemShut {NoStop}%
\bibitem [{\citenamefont {Boussarie}\ \emph {et~al.}(2023)\citenamefont {Boussarie} \emph {et~al.}}]{Boussarie:2023izj}%
  \BibitemOpen
  \bibfield  {author} {\bibinfo {author} {\bibfnamefont {R.}~\bibnamefont {Boussarie}} \emph {et~al.},\ }\href@noop {} {\  (\bibinfo {year} {2023})},\ \Eprint {http://arxiv.org/abs/2304.03302} {arXiv:2304.03302 [hep-ph]} \BibitemShut {NoStop}%
\bibitem [{\citenamefont {Pisano}\ \emph {et~al.}(2013)\citenamefont {Pisano}, \citenamefont {Boer}, \citenamefont {Brodsky}, \citenamefont {Buffing},\ and\ \citenamefont {Mulders}}]{Pisano:2013cya}%
  \BibitemOpen
  \bibfield  {author} {\bibinfo {author} {\bibfnamefont {C.}~\bibnamefont {Pisano}}, \bibinfo {author} {\bibfnamefont {D.}~\bibnamefont {Boer}}, \bibinfo {author} {\bibfnamefont {S.~J.}\ \bibnamefont {Brodsky}}, \bibinfo {author} {\bibfnamefont {M.~G.~A.}\ \bibnamefont {Buffing}}, \ and\ \bibinfo {author} {\bibfnamefont {P.~J.}\ \bibnamefont {Mulders}},\ }\href {\doibase 10.1007/JHEP10(2013)024} {\bibfield  {journal} {\bibinfo  {journal} {JHEP}\ }\textbf {\bibinfo {volume} {10}},\ \bibinfo {pages} {024} (\bibinfo {year} {2013})},\ \Eprint {http://arxiv.org/abs/1307.3417} {arXiv:1307.3417 [hep-ph]} \BibitemShut {NoStop}%
\bibitem [{\citenamefont {Jia}\ \emph {et~al.}(2024)\citenamefont {Jia}, \citenamefont {Zhou},\ and\ \citenamefont {Zhou}}]{Jia:2024xzx}%
  \BibitemOpen
  \bibfield  {author} {\bibinfo {author} {\bibfnamefont {Y.}~\bibnamefont {Jia}}, \bibinfo {author} {\bibfnamefont {J.}~\bibnamefont {Zhou}}, \ and\ \bibinfo {author} {\bibfnamefont {Y.-j.}\ \bibnamefont {Zhou}},\ }\href@noop {} {\  (\bibinfo {year} {2024})},\ \Eprint {http://arxiv.org/abs/2406.09381} {arXiv:2406.09381 [hep-ph]} \BibitemShut {NoStop}%
\bibitem [{\citenamefont {Wang}\ \emph {et~al.}(2024)\citenamefont {Wang}, \citenamefont {Wen}, \citenamefont {Xing},\ and\ \citenamefont {Yan}}]{Wang:2024zns}%
  \BibitemOpen
  \bibfield  {author} {\bibinfo {author} {\bibfnamefont {H.-L.}\ \bibnamefont {Wang}}, \bibinfo {author} {\bibfnamefont {X.-K.}\ \bibnamefont {Wen}}, \bibinfo {author} {\bibfnamefont {H.}~\bibnamefont {Xing}}, \ and\ \bibinfo {author} {\bibfnamefont {B.}~\bibnamefont {Yan}},\ }\href {\doibase 10.1103/PhysRevD.109.095025} {\bibfield  {journal} {\bibinfo  {journal} {Phys. Rev. D}\ }\textbf {\bibinfo {volume} {109}},\ \bibinfo {pages} {095025} (\bibinfo {year} {2024})},\ \Eprint {http://arxiv.org/abs/2401.08419} {arXiv:2401.08419 [hep-ph]} \BibitemShut {NoStop}%
\bibitem [{\citenamefont {Wen}\ \emph {et~al.}(2024{\natexlab{a}})\citenamefont {Wen}, \citenamefont {Yan}, \citenamefont {Yu},\ and\ \citenamefont {Yuan}}]{Wen:2024cfu}%
  \BibitemOpen
  \bibfield  {author} {\bibinfo {author} {\bibfnamefont {X.-K.}\ \bibnamefont {Wen}}, \bibinfo {author} {\bibfnamefont {B.}~\bibnamefont {Yan}}, \bibinfo {author} {\bibfnamefont {Z.}~\bibnamefont {Yu}}, \ and\ \bibinfo {author} {\bibfnamefont {C.~P.}\ \bibnamefont {Yuan}},\ }\href@noop {} {\  (\bibinfo {year} {2024}{\natexlab{a}})},\ \Eprint {http://arxiv.org/abs/2408.07255} {arXiv:2408.07255 [hep-ph]} \BibitemShut {NoStop}%
\bibitem [{\citenamefont {Wen}\ \emph {et~al.}(2024{\natexlab{b}})\citenamefont {Wen}, \citenamefont {Yan}, \citenamefont {Yu},\ and\ \citenamefont {Yuan}}]{Wen:2024nff}%
  \BibitemOpen
  \bibfield  {author} {\bibinfo {author} {\bibfnamefont {X.-K.}\ \bibnamefont {Wen}}, \bibinfo {author} {\bibfnamefont {B.}~\bibnamefont {Yan}}, \bibinfo {author} {\bibfnamefont {Z.}~\bibnamefont {Yu}}, \ and\ \bibinfo {author} {\bibfnamefont {C.~P.}\ \bibnamefont {Yuan}},\ }\href@noop {} {\  (\bibinfo {year} {2024}{\natexlab{b}})},\ \Eprint {http://arxiv.org/abs/2411.13845} {arXiv:2411.13845 [hep-ph]} \BibitemShut {NoStop}%
\bibitem [{\citenamefont {Cheng}\ and\ \citenamefont {Yan}(2025)}]{Cheng:2025cuv}%
  \BibitemOpen
  \bibfield  {author} {\bibinfo {author} {\bibfnamefont {K.}~\bibnamefont {Cheng}}\ and\ \bibinfo {author} {\bibfnamefont {B.}~\bibnamefont {Yan}},\ }\href@noop {} {\  (\bibinfo {year} {2025})},\ \Eprint {http://arxiv.org/abs/2501.03321} {arXiv:2501.03321 [hep-ph]} \BibitemShut {NoStop}%
\bibitem [{\citenamefont {Hatta}\ \emph {et~al.}(2021)\citenamefont {Hatta}, \citenamefont {Xiao}, \citenamefont {Yuan},\ and\ \citenamefont {Zhou}}]{Hatta:2021jcd}%
  \BibitemOpen
  \bibfield  {author} {\bibinfo {author} {\bibfnamefont {Y.}~\bibnamefont {Hatta}}, \bibinfo {author} {\bibfnamefont {B.-W.}\ \bibnamefont {Xiao}}, \bibinfo {author} {\bibfnamefont {F.}~\bibnamefont {Yuan}}, \ and\ \bibinfo {author} {\bibfnamefont {J.}~\bibnamefont {Zhou}},\ }\href {\doibase 10.1103/PhysRevD.104.054037} {\bibfield  {journal} {\bibinfo  {journal} {Phys. Rev. D}\ }\textbf {\bibinfo {volume} {104}},\ \bibinfo {pages} {054037} (\bibinfo {year} {2021})},\ \Eprint {http://arxiv.org/abs/2106.05307} {arXiv:2106.05307 [hep-ph]} \BibitemShut {NoStop}%
\bibitem [{\citenamefont {Shao}\ \emph {et~al.}(2023{\natexlab{b}})\citenamefont {Shao}, \citenamefont {Zhang}, \citenamefont {Zhou},\ and\ \citenamefont {Zhou}}]{Shao:2022stc}%
  \BibitemOpen
  \bibfield  {author} {\bibinfo {author} {\bibfnamefont {D.~Y.}\ \bibnamefont {Shao}}, \bibinfo {author} {\bibfnamefont {C.}~\bibnamefont {Zhang}}, \bibinfo {author} {\bibfnamefont {J.}~\bibnamefont {Zhou}}, \ and\ \bibinfo {author} {\bibfnamefont {Y.-J.}\ \bibnamefont {Zhou}},\ }\href {\doibase 10.1103/PhysRevD.107.036020} {\bibfield  {journal} {\bibinfo  {journal} {Phys. Rev. D}\ }\textbf {\bibinfo {volume} {107}},\ \bibinfo {pages} {036020} (\bibinfo {year} {2023}{\natexlab{b}})},\ \Eprint {http://arxiv.org/abs/2212.05775} {arXiv:2212.05775 [hep-ph]} \BibitemShut {NoStop}%
\bibitem [{\citenamefont {Shao}\ \emph {et~al.}(2023{\natexlab{c}})\citenamefont {Shao}, \citenamefont {Zhang}, \citenamefont {Zhou},\ and\ \citenamefont {Zhou}}]{Shao:2023zge}%
  \BibitemOpen
  \bibfield  {author} {\bibinfo {author} {\bibfnamefont {D.~Y.}\ \bibnamefont {Shao}}, \bibinfo {author} {\bibfnamefont {C.}~\bibnamefont {Zhang}}, \bibinfo {author} {\bibfnamefont {J.}~\bibnamefont {Zhou}}, \ and\ \bibinfo {author} {\bibfnamefont {Y.-j.}\ \bibnamefont {Zhou}},\ }\href {\doibase 10.1103/PhysRevD.108.116015} {\bibfield  {journal} {\bibinfo  {journal} {Phys. Rev. D}\ }\textbf {\bibinfo {volume} {108}},\ \bibinfo {pages} {116015} (\bibinfo {year} {2023}{\natexlab{c}})},\ \Eprint {http://arxiv.org/abs/2306.02337} {arXiv:2306.02337 [hep-ph]} \BibitemShut {NoStop}%
\bibitem [{\citenamefont {Davidson}\ \emph {et~al.}(2012)\citenamefont {Davidson}, \citenamefont {Nanava}, \citenamefont {Przedzinski}, \citenamefont {Richter-Was},\ and\ \citenamefont {Was}}]{Davidson:2010rw}%
  \BibitemOpen
  \bibfield  {author} {\bibinfo {author} {\bibfnamefont {N.}~\bibnamefont {Davidson}}, \bibinfo {author} {\bibfnamefont {G.}~\bibnamefont {Nanava}}, \bibinfo {author} {\bibfnamefont {T.}~\bibnamefont {Przedzinski}}, \bibinfo {author} {\bibfnamefont {E.}~\bibnamefont {Richter-Was}}, \ and\ \bibinfo {author} {\bibfnamefont {Z.}~\bibnamefont {Was}},\ }\href {\doibase 10.1016/j.cpc.2011.12.009} {\bibfield  {journal} {\bibinfo  {journal} {Comput. Phys. Commun.}\ }\textbf {\bibinfo {volume} {183}},\ \bibinfo {pages} {821} (\bibinfo {year} {2012})},\ \Eprint {http://arxiv.org/abs/1002.0543} {arXiv:1002.0543 [hep-ph]} \BibitemShut {NoStop}%
\bibitem [{\citenamefont {Bodrov}(2024)}]{Bodrov:2024wrw}%
  \BibitemOpen
  \bibfield  {author} {\bibinfo {author} {\bibfnamefont {D.}~\bibnamefont {Bodrov}},\ }\href {\doibase 10.1142/S0217751X24420065} {\bibfield  {journal} {\bibinfo  {journal} {Int. J. Mod. Phys. A}\ }\textbf {\bibinfo {volume} {39}},\ \bibinfo {pages} {2442006} (\bibinfo {year} {2024})},\ \Eprint {http://arxiv.org/abs/2405.16512} {arXiv:2405.16512 [hep-ex]} \BibitemShut {NoStop}%
\bibitem [{Bel(2020)}]{BelleII:2020tracking}%
  \BibitemOpen
  \href {https://docs.belle2.org/pub_data/documents/728/} {\bibfield  {journal} {\bibinfo  {journal} {BELLE2-NOTE-PL-2020-014}\ } (\bibinfo {year} {2020})}\BibitemShut {NoStop}%
\bibitem [{\citenamefont {Banerjee}(2022)}]{Banerjee:2022vdd}%
  \BibitemOpen
  \bibfield  {author} {\bibinfo {author} {\bibfnamefont {S.}~\bibnamefont {Banerjee}},\ }\href {\doibase 10.3390/universe8090480} {\bibfield  {journal} {\bibinfo  {journal} {Universe}\ }\textbf {\bibinfo {volume} {8}},\ \bibinfo {pages} {480} (\bibinfo {year} {2022})},\ \Eprint {http://arxiv.org/abs/2209.11639} {arXiv:2209.11639 [hep-ex]} \BibitemShut {NoStop}%
\bibitem [{\citenamefont {Atwood}\ and\ \citenamefont {Soni}(1992)}]{Atwood:1991ka}%
  \BibitemOpen
  \bibfield  {author} {\bibinfo {author} {\bibfnamefont {D.}~\bibnamefont {Atwood}}\ and\ \bibinfo {author} {\bibfnamefont {A.}~\bibnamefont {Soni}},\ }\href {\doibase 10.1103/PhysRevD.45.2405} {\bibfield  {journal} {\bibinfo  {journal} {Phys. Rev. D}\ }\textbf {\bibinfo {volume} {45}},\ \bibinfo {pages} {2405} (\bibinfo {year} {1992})}\BibitemShut {NoStop}%
\bibitem [{\citenamefont {Inami}\ \emph {et~al.}(2003)\citenamefont {Inami} \emph {et~al.}}]{Belle:2002nla}%
  \BibitemOpen
  \bibfield  {author} {\bibinfo {author} {\bibfnamefont {K.}~\bibnamefont {Inami}} \emph {et~al.} (\bibinfo {collaboration} {Belle}),\ }\href {\doibase 10.1016/S0370-2693(02)02984-2} {\bibfield  {journal} {\bibinfo  {journal} {Phys. Lett. B}\ }\textbf {\bibinfo {volume} {551}},\ \bibinfo {pages} {16} (\bibinfo {year} {2003})},\ \Eprint {http://arxiv.org/abs/hep-ex/0210066} {arXiv:hep-ex/0210066} \BibitemShut {NoStop}%
\bibitem [{\citenamefont {Huang}\ \emph {et~al.}(1997)\citenamefont {Huang}, \citenamefont {Lu},\ and\ \citenamefont {Tao}}]{Huang:1996jr}%
  \BibitemOpen
  \bibfield  {author} {\bibinfo {author} {\bibfnamefont {T.}~\bibnamefont {Huang}}, \bibinfo {author} {\bibfnamefont {W.}~\bibnamefont {Lu}}, \ and\ \bibinfo {author} {\bibfnamefont {Z.-j.}\ \bibnamefont {Tao}},\ }\href {\doibase 10.1103/PhysRevD.55.1643} {\bibfield  {journal} {\bibinfo  {journal} {Phys. Rev. D}\ }\textbf {\bibinfo {volume} {55}},\ \bibinfo {pages} {1643} (\bibinfo {year} {1997})},\ \Eprint {http://arxiv.org/abs/hep-ph/9609220} {arXiv:hep-ph/9609220} \BibitemShut {NoStop}%
\bibitem [{\citenamefont {Bernreuther}\ \emph {et~al.}(1997)\citenamefont {Bernreuther}, \citenamefont {Brandenburg},\ and\ \citenamefont {Overmann}}]{Bernreuther:1996dr}%
  \BibitemOpen
  \bibfield  {author} {\bibinfo {author} {\bibfnamefont {W.}~\bibnamefont {Bernreuther}}, \bibinfo {author} {\bibfnamefont {A.}~\bibnamefont {Brandenburg}}, \ and\ \bibinfo {author} {\bibfnamefont {P.}~\bibnamefont {Overmann}},\ }\href {\doibase 10.1016/S0370-2693(96)01501-8} {\bibfield  {journal} {\bibinfo  {journal} {Phys. Lett. B}\ }\textbf {\bibinfo {volume} {391}},\ \bibinfo {pages} {413} (\bibinfo {year} {1997})},\ \bibinfo {note} {[Erratum: Phys.Lett.B 412, 425--425 (1997)]},\ \Eprint {http://arxiv.org/abs/hep-ph/9608364} {arXiv:hep-ph/9608364} \BibitemShut {NoStop}%
\bibitem [{\citenamefont {Bernreuther}\ \emph {et~al.}(2021)\citenamefont {Bernreuther}, \citenamefont {Chen},\ and\ \citenamefont {Nachtmann}}]{Bernreuther:2021elu}%
  \BibitemOpen
  \bibfield  {author} {\bibinfo {author} {\bibfnamefont {W.}~\bibnamefont {Bernreuther}}, \bibinfo {author} {\bibfnamefont {L.}~\bibnamefont {Chen}}, \ and\ \bibinfo {author} {\bibfnamefont {O.}~\bibnamefont {Nachtmann}},\ }\href {\doibase 10.1103/PhysRevD.103.096011} {\bibfield  {journal} {\bibinfo  {journal} {Phys. Rev. D}\ }\textbf {\bibinfo {volume} {103}},\ \bibinfo {pages} {096011} (\bibinfo {year} {2021})},\ \Eprint {http://arxiv.org/abs/2101.08071} {arXiv:2101.08071 [hep-ph]} \BibitemShut {NoStop}%
\end{thebibliography}%
\end{document}